\documentclass[12pt]{article}
\usepackage{a4wide}

\usepackage[dvipdfmx,hiresbb]{graphicx} 
\usepackage{mediabb}                    

\usepackage{amsmath,amssymb}
\usepackage{color}
\usepackage[bookmarks,bookmarksnumbered]{hyperref}
\usepackage{cellspace}
\usepackage{tikz}
\usepackage{mathtools}
\usepackage{cite}
\usetikzlibrary{fadings}
\usetikzlibrary{patterns}
\usetikzlibrary{shadows.blur}
\usetikzlibrary{shapes}
\newcommand{\aln}[1]{\begin{align}#1\end{align}}
\newcommand{\nn}{\nonumber\\}

\begin{document}
\title{\vspace{-3cm}
\vbox{
\baselineskip 14pt
\hfill \hbox{\normalsize KEK-TH-2557
}}  \vskip 1cm
\bf \Large Effective Brane Field Theory with \\ Higher-form Symmetry 
\vskip 0.5cm
}
\author{
Yoshimasa Hidaka\thanks{E-mail: \tt hidaka@post.kek.jp}\, and
Kiyoharu Kawana\thanks{E-mail: \tt kkiyoharu@kias.re.kr}
\bigskip\\
\normalsize
\it $^*$ KEK Theory Center, Tsukuba 305-0801, Japan\\
\normalsize
\it $^*$ Graduate Institute for Advanced Studies, SOKENDAI, Tsukuba 305-0801, Japan\\
\normalsize
\it $^*$ Department of Physics, The University of Tokyo,\\
\normalsize
\it  7-3-1 Hongo, Bunkyo-ku, Tokyo  113-0033, Japan\\
\normalsize
\it $^*$ International Center for Quantum-field Measurement Systems\\
\normalsize
\it for Studies of the Universe and Particles (QUP), KEK, Tsukuba, 305-0801, Japan\\
\normalsize
\it $^*$ RIKEN iTHEMS, RIKEN, Wako 351-0198, Japan\\
\normalsize
\it $^\dagger$ School of Physics, Korean Institute for Advanced Study, Seoul 02455, Korea
\smallskip
}
\date{\today}

\maketitle   
\begin{abstract} 
We propose an effective field theory for branes with higher-form symmetry as a generalization of ordinary Landau theory, 
%
which is an extension of the previous work by Iqbal and McGreevy for one-dimensional objects to an effective theory for $p$-dimensional objects. 
In the case of a $p$-form symmetry, the fundamental field $\psi[C_p^{}]$ is a functional of $p$-dimensional closed brane $C_p^{}$ embedded in a spacetime. 
As a natural generalization of ordinary field theory, we call this theory the {\it brane field theory}. 
In order to construct an action that is invariant under higher-form transformation, we generalize the idea of {\it area derivative} for one-dimensional objects to higher-dimensional ones. 
Following this, we discuss various fundamental properties of the brane field based on the higher-form invariant action.    
It is shown that the classical solution exhibits the area law in the unbroken phase of $\mathrm{U}(1)$ $p$-form symmetry, 
while it indicates a constant behavior in the broken phase for the large volume limit of $C_p^{}$.   
In the latter case, the low-energy effective theory is described by the $p$-form Maxwell theory.    
We also discuss brane-field theories with a discrete higher-form symmetry and show that the low-energy effective theory becomes a BF-type topological field theory, resulting in topological order.
%
Finally, we present a concrete brane-field model that describes a superconductor from the point of view of higher-form symmetry. 

\end{abstract} 

\setcounter{page}{1} 

\newpage  

\tableofcontents   

\newpage  

\section{Introduction}\label{Sec:intro}
Symmetry is one of the most important and fundamental concepts in modern physics, and it plays an essential role in classifying phases of vacuum and matter. 
For instance, various phase transitions can be comprehended through the presence of symmetries and their spontaneous breaking. 
The Landau theory provides a comprehensive and effective framework for this description~\cite{Landau:1937obd,landau2013statistical}. 
In the Landau theory and its extensions, an order parameter field $\phi(X)$, charged under a global symmetry, is introduced, and the theory (i.e., free energy, Hamiltonian, or Lagrangian) is constructed as invariant under the symmetry. 
Furthermore, it is important to note that the assumption of the conventional Landau theory is that the order parameter $\phi(X)$ is a local function of the spacetime point.
In this sense, the conventional Landau theory describes an effective theory of point-like or zero-dimensional objects, such as particles.

The concept of symmetry has been recently extended in various directions, called ``generalized global symmetries’’~\cite{Gaiotto:2014kfa}, which include higher-form~\cite{Kapustin:2005py,Pantev:2005zs,Nussinov:2009zz,Banks:2010zn,Kapustin:2013uxa,Aharony:2013hda,Kapustin:2014gua,Gaiotto:2017yup,Hirono:2018fjr,Hidaka:2019jtv,Hidaka:2022blq}, higher-group~\cite{
Sharpe:2015mja,Tachikawa:2017gyf,Cordova:2018cvg,Benini:2018reh,Tanizaki:2019rbk,DelZotto:2020sop,Hidaka:2020iaz,Hidaka:2020izy,Brennan:2020ehu,Hidaka:2021mml,Hidaka:2021kkf,Apruzzi:2021mlh,Barkeshli:2022edm,Nakajima:2022feg,Radenkovic:2022qkd,Bhardwaj:2022scy,Kan:2023yhz}, and noninvertible symmetries~\cite{Bhardwaj:2017xup,Chang:2018iay,Ji:2019jhk,Komargodski:2020mxz,Nguyen:2021yld,Heidenreich:2021xpr,Koide:2021zxj,Kaidi:2021xfk,Choi:2021kmx,Roumpedakis:2022aik,Bhardwaj:2022yxj,Cordova:2022ieu,Bashmakov:2022jtl,Choi:2022rfe,Bartsch:2022mpm,Apruzzi:2022rei,GarciaEtxebarria:2022vzq,Niro:2022ctq,Chen:2022cyw,Bashmakov:2022uek,Karasik:2022kkq,GarciaEtxebarria:2022jky,Choi:2022fgx,Yokokura:2022alv,Bhardwaj:2022maz,Bartsch:2022ytj,Kaidi:2023maf,Lin:2023uvm,Chen:2023czk}.
Like ordinary symmetries, these generalized symmetries are powerful tools, which can be applied to phenomena such as spontaneous breaking, anomalies, topological orders, and symmetry-protected topological phases (For recent reviews, lecture notes, and complementary references, see Refs.~\cite{McGreevy:2022oyu,Gomes:2023ahz,Schafer-Nameki:2023jdn,Brennan:2023mmt,Bhardwaj:2023kri,Luo:2023ive,Shao:2023gho}.)
A generalized symmetry consists of an algebra of symmetry operators, which includes the compositions and intersections.
The symmetry operators are topological, meaning that the deformation of symmetry operators in spacetime does not change the observables as long as they do not contact charged objects. 
Among these generalized symmetries, the most fundamental one is the higher $p$-form symmetry. 
This extends the ordinary symmetry from $0$-dimensional objects to $p$-dimensional extended objects, such as Wilson loops, domain walls, and vortices. 
The symmetry operators correspond to topological objects with codimension $(p+1)$ in spacetime. 
Compositions of these objects form a group, and the linking with $p$-dimensional charged objects in spacetime generates a symmetry transformation.

Considering the great success of the Landau theory, a natural question arises: Is it possible to construct effective field theories with extended objects for higher-form symmetries?
The purpose of this paper is to explore this possibility, and demonstrate that this framework provides an effective approach to understanding the physics of higher-form symmetry, much as the conventional Landau theory does for $0$-form symmetries.
The advantage of this approach is that, within the mean-field approximation, it naturally describes the phase transition to topological order by using extended order parameters, which cannot be explained in conventional Landau theory.
It should be mentioned that our approach is inspired by Ref.~\cite{Iqbal:2021rkn}, where the effective field theory for $1$-form symmetry is introduced as {\it mean string field theory}. 
Generalizing it, we refer to our field theory for $p$-form symmetry as {\it effective brane field theory}. 

To construct the brane field theory, we should clarify which type of $p$-dimensional branes $C_p^{}$ should be considered.
In this paper, we focus on $p$-dimensional closed branes $C_p^{}$ that extend spatially in a $d$-dimensional Lorentzian spacetime $\Sigma_d^{}$~\footnote{
In general, we can also consider $p$-dimensional objects that extend the time direction in $\Sigma_d^{}$.
The construction of the brane field theory is completely parallel to the spatially extended case, but with different signatures. 
In this paper, we focus on the spatially extended objects for simplicity.
}.
In other words, $C_p^{}$ is represented by the spacetime embedding $\{X^\mu(\xi)\}_{\mu=0}^{d-1}$, where $\xi=(\xi^1,\xi^2,\cdots,\xi^p)$ denotes the intrinsic coordinates.
Then, the brane field  $\psi[C_p^{}]$ is no longer just a function of spacetime point, but a functional of $\{X^\mu(\xi)\}_{\mu=0}^{d-1}$. 
If we allow any functional forms, there would be little hope that we can obtain controllable brane field theory even at the classical level.
Thus, it is natural to impose physically reasonable conditions as in the ordinary quantum field theory: spacetime diffeomorphism invariance and reparametrization invariance.
Namely, we assume that $\psi[C_p^{}]$ behaves as a scalar under these transformations.

Not only the concept of the field but also the concept of {\it derivative} must be generalized in order to construct a brane-field action invariant under higher-form transformations. 
In general, the variation of the brane field with respect to a small change of the subspace is described by the functional derivative $\delta \psi[C_p^{}]/\delta X^\mu(\xi)$. 
On a $p$-dimensional object $C_p^{}$, we can generally consider variations of subspaces of lower dimensions, and the functional derivative contains all such contributions. 
In this paper, however, we focus on a $p$-dimensional variation $\delta C_p^{}$ such that the corresponding functional derivative is described by the {\it area derivative}, which was originally introduced for one-dimensional objects~\cite{Migdal:1983qrz,Makeenko:1980vm,Polyakov:1980ca}.
%
%
We will see that, as the ordinary derivative for $p=0$ is given by the one-form $\partial_\mu^{}\phi(X)dX^\mu$, the area derivative on the $p$-dimensional subspace $C_p^{}$ is given by a $(p+1)$-form functional derivative as shown in Eq.~(\ref{area derivative form}).
In this sense, the area derivative can be interpreted as one of the natural generalizations of the ordinary derivative $\partial_\mu^{}\phi(X)$.

Following our discussion on the construction of the brane-field theory, we perform a mean-field analysis.
First, we show that the classical solution $\langle \psi[C_p^{}]\rangle$ exhibits the area-law behavior in the unbroken phase of $\mathrm{U}(1)$ $p$-form symmetry, while it is constant in the broken phase in the large volume limit of $C_p^{}$.
These behaviors can be naturally interpreted as a generalization of the off-diagonal-long-range order of the two-point correlation function $\langle \phi^\dagger (x)\phi(y)\rangle$ for the $0$-form symmetries.
Second, by considering phase fluctuations of the order parameter, we show that the low-energy effective theory in the broken phase of $\mathrm{U}(1)$ $p$-form symmetry is given by the $p$-form Maxwell theory, which is a $p$-form version of the Nambu-Goldstone theorem for $0$-form symmetries~\cite{Nambu:1961tp,Goldstone:1961eq,Goldstone:1962es,Kapustin:2014gua,Lake:2018dqm,Hofman:2018lfz}. 
Note that since we are considering theories for extended objects in spacetime, the effective theory can contain many local fluctuations other than the $p$-form gauge field as a Nambu-Goldstone field. 
However, these fields typically become massive because they are not protected by the $p$-form symmetry. 
We will explicitly show this for the spacetime scalar mode as an example. 
Third, as well as the $0$-form case, we can also consider discrete higher-form symmetry and its breaking in the present brane-field theory.
By generalizing the discussion of $0$-form symmetry in Ref.~\cite{Hidaka:2019mfm} to the $p$-form symmetry case, we derive the low-energy effective theory in the broken phase. This theory takes the form of a $\mathrm{BF}$-type topological field theory and exhibits topological order.
Finally, we discuss a concrete brane-field model for a superconductor and derive its low-energy effective theory in the superconducting (Higgs) phase.

The organization of this paper is as follows. 
In Sec.~\ref{brane field theory}, we introduce the $p$-brane field and the field theory with $\mathrm{U}(1)$ $p$-form symmetry, and discuss several technical aspects, including the generalization of the area derivatives and the construction of the Noether current. 
In Sec.~\ref{sec:SSB}, we focus on the spontaneous breaking of higher-form symmetry. 
Using the expectation value of the brane field as the order parameter, we discuss the spontaneous breaking of $p$-form symmetry within the mean-field approximation, the low-energy effective theory, and emergent symmetries in the broken phase. We show that the effective theories for the spontaneous breaking of continuous and discrete higher-form symmetries are the $p$-form Maxwell theory and the BF-type topological field theory, respectively.
We also discuss the brane field model for a superconductor and its effective theory.
Section~\ref{sec:summary} is devoted to summary and discussion. 
The Appendices provide additional details on differential forms, truncated action, and other related calculations.
%

\section{Brane field theory}\label{brane field theory}
We explain how to construct field theory for higher-dimensional branes $C_p^{}$. 
We first introduce the brane field $\psi[C_p^{}]$ by imposing two physically natural conditions: spacetime diffeomorphism invariance and reparametrization invariance.  
Then, we discuss the relation between the functional derivative and the ``{\it area derivative}''~\cite{Migdal:1983qrz,Makeenko:1980vm,Polyakov:1980ca}, which is a natural generalization of the ordinary derivative of a local field $\partial_\mu^{}\phi(x)$.    
%

\subsection{\texorpdfstring{$p$}{p}-brane field}\label{}
We discuss how to construct the brane field $\psi[C_p^{}]$. 
We consider a $d$-dimensional spacetime manifold $\Sigma_d$ with the metric $g_{\mu\nu}$.  
We employ the Minkowski metric signature for $d$-spacetime dimensions as $(-,+,+,\cdots)$.  
$C_p^{}$ is a subspace in $\Sigma_d$, which can be expressed by an embedding function $S^p\to \Sigma_d$, i.e., $\{X^\mu(\xi)\}_{\mu=0}^{d-1}$, where $S^p$ is a $p$-dimensional space.
Therefore, as mentioned in Introduction, $\psi[C_p^{}]$ can be thought of a functional of $\{X^\mu(\xi)\}_{\mu=0}^{d-1}$. 
Since we are interested in a brane as a $p$-dimensional object at a given time of some specific time choice, we will restrict 
$C_p$ to spacelike objects. 
$C_p$ may have a boundary; however, we mainly focus our discussion on the case where $C_p$ has no boundary. 

In general, $\psi[C_p^{}]$ can take any functionals of $\{X^\mu(\xi)\}_{\mu=0}^{d-1}$, but we restrict it by imposing the following conditions as in the ordinary field theory:
\begin{description}
\item [\bf Spacetime diffeomorphism]: $\psi[C_p^{}]$ is a scalar under the spacetime diffeomorphism $X^\mu\rightarrow X'^\mu=f^\mu(X)$:
\aln{
\psi'[C_p^{'}]=\psi[\{X'^\mu(\xi)\}]=\psi[\{X^\mu(\xi)\}]=\psi[C_p^{}]~.\label{eq:spacetimediffeo}
}
\item [\bf Reparametrization invariance]: $\psi[C_p^{}]$ is invariant under the reparametrization on $C_p^{}$: $\xi^i\rightarrow \xi'^i=g^i(\xi)$:
\aln{
\psi'[C_p^{'}]=\psi[\{X^\mu(\xi')\}]=\psi[\{X^\mu(\xi)\}]=\psi[C_p^{}]~.
} 
\end{description}
We note that we imposed a scalar condition \eqref{eq:spacetimediffeo} on the spacetime diffeomorphism to simplify the argument; more generally, it could take a covariant form. 
Typical examples that satisfy the above conditions are the functionals of various differential forms:
\aln{
\psi[C_p^{}]=\psi\left(\left\{\int_{C_p^{}} d^p\xi \sqrt{h}A^{(a)}(X(\xi))\right\}\right)~,\label{functional form}
} 
where $A^{(a)}(X)$ is a spacetime scalar, and $h=\det(h_{ij}^{})$ is the determinant of the induced metric,
\aln{
h_{ij}^{}(\xi)=\frac{\partial X^\mu(\xi)}{\partial \xi^i}\frac{\partial X^\nu(\xi)}{\partial \xi^j}g_{\mu\nu}^{}(X(\xi))=e^{\mu}_i(\xi) e^{\nu}_j(\xi) g_{\mu\nu}^{}~,\quad \det(h_{ij}^{})>0~.
\label{induced metric}
}
Besides, the index $a$ represents various types of volume integrals. 
Note that for a given scalar $A^{(a)}(X)$ we can always rewrite the volume integral in Eq.~(\ref{functional form}) by a $p$-form integration as 
\aln{
\int d^p\xi \sqrt{h}A^{(a)}(X(\xi))=\int_{C_p^{}} A_p^{(a)}~,
\label{p-form integration}
}
where $A_p^{(a)}=\frac{1}{p!}A^{(a)}_{\mu_1^{}\cdots \mu_{p}^{}}(X(\xi))dX^{\mu_1^{}}\wedge \cdots \wedge dX^{\mu_p^{}}$ is the $p$-form satisfying 
\aln{\frac{1}{p!}\eta^{i_1^{}\cdots i_p^{}}e_{i_1^{}}^{\mu_1^{}}(\xi)\cdots e_{i_p^{}}^{\mu_p^{}}(\xi)A^{(a)}_{\mu_1^{}\cdots \mu_{p}^{}}(X(\xi))=A^{(a)}(X(\xi))~. 
}
Here, $\eta^{j_1^{}\cdots j_p^{}}$ is the totally anti-symmetric tensor defined by Eq.~(\ref{totally anti-symmetric tensor}) in Appendix~\ref{differential forms}.   
$A^{(a)}_{\mu_1^{}\cdots \mu_{p}^{}}(X(\xi))$ can be explicitly written as  
\aln{
A^{(a)}_{\mu_1^{}\cdots \mu_{p}^{}}(X(\xi))
&=g_{\mu_1^{}\nu_1^{}}\cdots g_{\mu_p^{}\nu_p^{}}E^{\nu_1^{}\cdots \nu_p^{}}(\xi)A^{(a)}(X(\xi))
\notag\\
&=E_{\mu_1^{}\cdots \mu_p^{}}^{}(X(\xi))A^{(a)}(X(\xi))~,
\label{relation to p-form}
}
where
\aln{E^{\mu_1^{}\cdots \mu_p^{}}(X(\xi))\coloneqq\eta^{i_1^{}\cdots i_p^{}}e_{i_1^{}}^{\mu_1^{}}(\xi)\cdots e_{i_p^{}}^{\mu_p^{}}(\xi)~.
}
By introducing a $p$-form, 
\aln{E_p^{}\coloneqq\frac{1}{p!}E_{\mu_1^{}\cdots \mu_p^{}}(X(\xi))dX^{\mu_1^{}}\wedge \cdots \wedge dX^{\mu_p^{}}~,\label{Ep form}
}
we can also check 
\aln{E_{\mu_1^{}\cdots \mu_p^{}}^{}E^{\mu_1^{}\cdots \mu_p^{}}=p!\quad \leftrightarrow \quad E_p^{}\wedge \star E_p^{}=\sqrt{-g} dX^0\wedge\cdots \wedge dX^{d-1}~,
\label{EE}
}
and 
\aln{
\mathrm{Vol}[C_p^{}]=\int_{C_p^{}}E_p^{}~,
}
where $\mathrm{Vol}[C_p^{}]$ is the volume of $C_p$.

\subsection{Functional derivative and Area derivative}\label{Sec:volume integral}
In general, a variation of the brane field $\psi[C_p^{}]$ for an arbitrary change of the manifold $\delta C_p^{}=\{\delta X^{\mu}(\xi)\}_{\mu=0}^{d-1}$ is given by the functional derivative as
\aln{\label{general functional derivative}
\delta \psi[C_p^{}]=\int d^p\xi\sqrt{h}~\delta X^\mu(\xi)\frac{\delta \psi[C_p^{}]}{\delta X^\mu(\xi)}~.
}
\begin{figure}
    \centering
    \includegraphics[scale=0.4]{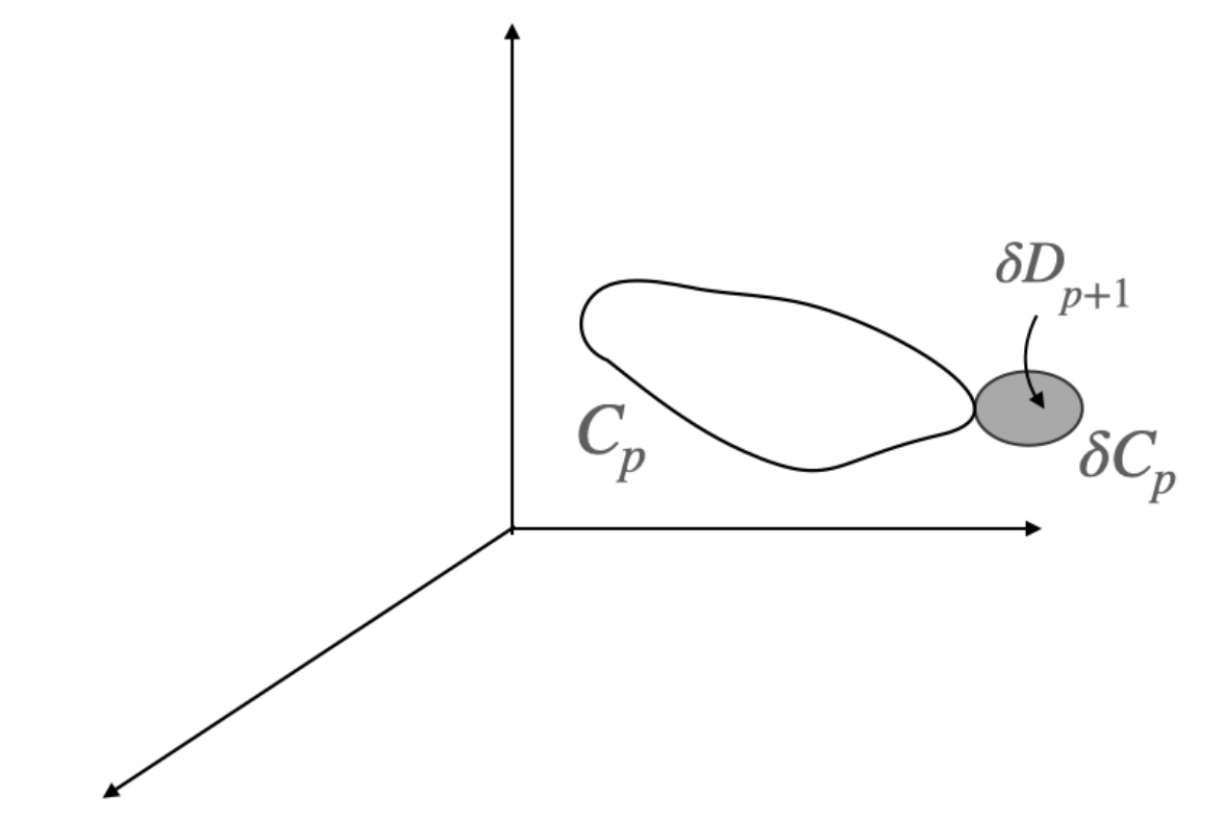}
    \caption{A $(p+1)$-dimensional subspace $\delta D_{p+1}^{}$ with a boundary $\delta C_p^{}$.
    }
    \label{fig:Dp+1}
\end{figure}
In particular, when $\delta C_p^{}$ has a small support around some point $\xi$ and is given by the boundary of a $(p+1)$-dimensional subspace, i.e., $ ^\exists \delta D_{p+1}^{},~\partial \delta D_{p+1}^{}=\delta C_p^{}$ (see Fig.~\ref{fig:Dp+1} as an example), Eq.~(\ref{general functional derivative}) can be also written as
\aln{\label{def:area derivative}
\delta \psi[C_p^{}]=\frac{1}{(p+1)!}\sigma^{\mu_1^{}\cdots \mu_{p+1}^{}}(\delta C_p^{})\frac{\delta \psi[C_p^{}]}{\delta \sigma^{\mu_1^{}\cdots \mu_{p+1}^{}}(\xi)}~,
}
where\footnote{When $p=0$, $\sigma^{\mu_1^{}\cdots \mu_{p+1}^{}}(\delta C_p^{})$ is just $\delta X^\mu$. 
}
\aln{
\sigma^{\mu_1^{}\cdots \mu_{p+1}^{}}(\delta C_p^{})&=
\int_{\delta D_{p+1}^{}}dX^{\mu_1^{}}\wedge \cdots \wedge dX^{\mu_{p+1}^{}}~
\notag 
\\
&=\int_{\delta D_{p+1}^{}}d(\delta X^{[\mu_1^{}}\wedge \cdots \wedge dX^{\mu_{p+1}^{}]})~
\notag \\
&=
\int_{\delta C_{p}^{}} \delta X^{[\mu_{1}^{}}dX^{\mu_2^{}}\wedge\cdots \wedge dX^{\mu_{p+1}^{}]}~.
\label{sigma volume}
}
Here, we have used the Stokes theorem in the third line of Eq.~\eqref{sigma volume}, and introduced the antisymmetrization,
\aln{
A^{[\mu_1^{}\mu_2^{}\cdots \mu_p^{}]}:=\frac{1}{p!}\sum_{\sigma \in S_{p}^{}}\mathrm{sgn}(\sigma)A^{\sigma(\mu_1^{})\sigma(\mu_2^{})\cdots \sigma(\mu_p^{})}~,
\label{eq:antisymmetrization}
}
where $S_p^{}$ is the symmetric group of degree $p$, and $\mathrm{sgn}(\sigma)$ is the sign of permutations. 
We call $\delta\psi[C_p^{}]/\delta \sigma^{\mu_1^{}\cdots \mu_{p+1}^{}}(\xi)$ the $p$-th order {\it area derivative}, 
which is the generalization of the area derivative for one-dimensional objects~\cite{Makeenko:1980vm,Polyakov:1980ca,Migdal:1983qrz,Iqbal:2021rkn} to higher-dimensional ones. 
The definition of the are derivative (\ref{def:area derivative}) is abstract, and we should clarify its relation to the ordinary functional derivative (\ref{general functional derivative}).   
Equation~(\ref{sigma volume}) can be more explicitly written as
\aln{
\sigma^{\mu_1^{}\cdots \mu_{p+1}^{}}(\delta C_p^{})
&=
\int_{\delta C_{p}^{}}d^p\xi\sqrt{h} 
\delta X^{[\mu_{1}^{}}(\xi) E^{\mu_2^{}\cdots \mu_{p+1}^{}]}(\xi)~,
}  
which leads to the following expression of $\delta \psi[C_p^{}]$:
\aln{
\delta \psi[C_p^{}]=
\frac{1}{(p+1)!}
\int_{\delta C_{p}^{}}d^p\xi \sqrt{h}
\delta X^{[\mu_{1}^{}}(\xi)E^{\mu_2^{}\cdots \mu_{p+1}^{}]}(\xi)
\frac{\delta \psi[C_p^{}]}{\delta \sigma^{\mu_1^{}\cdots \mu_{p+1}^{}}(\xi)}~.
}
This coincides with Eq.~(\ref{general functional derivative}) by the identification 
\aln{
\frac{\delta \psi[C_p^{}]}{\delta X^{\mu_{1}^{}}(\xi)}=
\frac{1}{(p+1)!}
E^{\mu_{2}^{}\cdots \mu_p+1^{}}(\xi)
\frac{\delta \psi[C_p^{}]}{\delta \sigma^{\mu_1^{}\cdots \mu_{p+1}^{}}(\xi)}~.
}
Note that for more general variation $\delta C_p^{}$, the above relation does not necessarily hold, and additional terms can appear on the right-hand side\footnote{For example, for $p=1$, 
there is also another term called the {\it path derivative} which corresponds to the variation $\delta C_1^{}$ such that an infinitesimal path $\delta\Gamma$ is added to a point $\{X^\mu(\xi_0^{})\}$ of the original loop. 
In this case, the functional derivative becomes~\cite{Makeenko:2002uj} $\delta /\delta X^\mu(\xi)=\partial_\mu^{}|_{X=X(\xi_0^{})}\delta(\xi-\xi_0^{})$. 
As a result, the functional derivative is given by the sum of the area derivative and path derivative for $p=1$. 
}.

In particular, the area derivative of the volume integral of a differential $p$ form $A_p^{(a)}$ can be calculated in the following way.     
Under the infinitesimal change $C_p^{}+\delta C_p^{}$, the variation of Eq.~(\ref{p-form integration}) is 
\aln{\label{p form variation}
\int_{\delta C_p^{}}A_p^{(a)}=\int_{\delta D_{p+1}^{}}dA_p^{(a)}&=\frac{1}{(p+1)!}\int_{\delta D_{p+1}^{}}dX^{\mu_p^{}}\wedge \cdots \wedge dX^{\mu_{p+1}^{}}F_{\mu_1^{}\cdots \mu_{p+1}^{}}^{(a)}(X)
\notag\\
&=\frac{1}{(p+1)!}\sigma^{\mu_1^{}\cdots \mu_{p+1}^{}}(\delta C_p^{})F_{\mu_1^{}\cdots \mu_{p+1}^{}}^{(a)}(X)~,
}
where 
\aln{
F_{p+1}^{(a)}=dA_p^{(a)}=\frac{1}{(p+1)!}F_{\mu_1^{}\cdots \mu_{p+1}^{}}^{(a)}(X)dX^{\mu_1^{}}\wedge \cdots \wedge dX^{\mu_{p+1}^{}}~.
} 
The above equation implies 
\aln{
\frac{\delta}{\delta \sigma^{\mu_1^{}\cdots \mu_{p+1}^{}}(\xi)}\left(\int_{C_p^{}}A_p^{(a)}\right)=F_{\mu_1^{}\cdots \mu_{p+1}^{}}^{(a)}(X)~.
}
Thus, as long as we consider the brane field whose functional form is given by Eq.~(\ref{functional form}), 
the area derivative is given by
\aln{
\frac{\delta \psi[C_p^{}]}{\delta \sigma^{\mu_1^{}\cdots \mu_{p+1}^{}}(\xi)}=\sum_a F^{(a)}_{\mu_1^{}\cdots \mu_{p+1}^{}}(X(\xi))\frac{\partial \psi(\{z^a\})}{\partial z^a}\bigg|_{z^a=\int_{C_p^{}}A_p^{(a)}}~.
\label{general area derivative}
}  
It is also convenient to introduce the $(p+1)$-form of the area derivative:
\aln{\label{area derivative form}
D\psi[C_p^{}]\coloneqq\frac{1}{(p+1)!}\frac{\delta \psi[C_p^{}]}{\delta \sigma^{\mu_1^{}\cdots\mu_{p+1}^{}}(\xi)}dX^{\mu_1^{}}\wedge \cdots \wedge dX^{\mu_{p+1}^{}}~.
}
Let us see a few important examples below.  

\

\noindent {\bf Wilson surface}\\
A first important but trivial example is the Wilson surface defined by
\aln{\psi[C_p^{}]=W[C_p^{}]\coloneqq\exp\left(i\int_{C_p^{}}A_p^{}\right)~,
}
where $A_p^{}$ is a $p$-form field. 
As already seen before, the area derivative is given by
\aln{
DW[C_p^{}]=iW[C_p^{}]F_{p+1}^{}~,
\label{Wilson loop are derivative}
}
where $F_{p+1}^{}=dA_p^{}$ is the field strength.

\

\noindent {\bf World volume}\\
The next important example is the world volume, 
\aln{\psi[C_p^{}]=\mathrm{Vol}[C_p^{}]=\int d^p\xi\sqrt{h}~.
\label{world volume}
}
This case corresponds to $A^{(a)}(X(\xi))=1$ in Eq.~(\ref{functional form}). 
Thus, by using Eq.~(\ref{relation to p-form}), we have
\aln{
D\mathrm{Vol}[C_p^{}]=dE_p^{}~.
}

\

\noindent {\bf Minimal volume}\\
Another example is the (minimal) volume $\mathrm{Vol}[M_{p+1}^{}]$ of a $(p+1)$-dimensional subspace $M_{p+1}^{}$ enclosed by $C_p^{}$, i.e., $\partial M_{p+1}^{}=C_p^{}$. 
\begin{figure}
    \centering
    \includegraphics[scale=0.4]{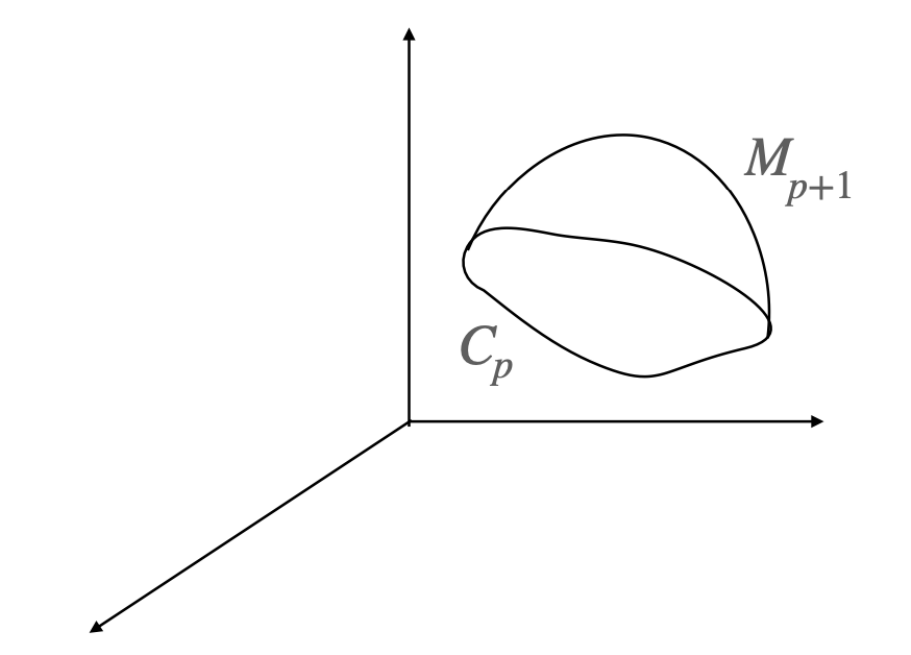}
    \caption{A $(p+1)$-dimensional subspace $M_{p+1}^{}$ with a boundary $C_p^{}$. 
    }
    \label{fig:Mp+1}
\end{figure}
See Fig.~\ref{fig:Mp+1} as an example. 
In this case, we only regard the boundary subspace $C_p^{}$ as a physical variable. 
Since $\mathrm{Vol}[M_{p+1}^{}]$ is given by the $(p+1)$-form integral on $M_{p+1}^{}$, this case does not belong to Eq.~(\ref{general area derivative}), but we can calculate its area derivative as follows. 
A general variation $\delta C_p^{}$ can be constructed by adding infinitesimal small loop $\delta C_p^{\rm Loop}$ at each point on $C_p^{}$. 
By representing the bulk of $\delta \Gamma_p^{}$ as $\delta D_{p+1}^{}$, we have
\aln{
\delta \mathrm{Vol}[M_{p+1}^{}]=\int_{\delta D_{p+1}^{}}E_{p+1}^{}~,
}
which corresponds to the expression~(\ref{p form variation}).
Thus, one can see that $E_{p+1}^{}$ corresponds to $F_{p+1}^{}$ in this case, and we have 
\aln{
D\mathrm{Vol}[M_{p+1}^{}]=
E_{p+1}^{}~.\label{area derivative of volume}
}

\subsection{Brane field action}
Now, we define the brane field action with global $\mathrm{U}(1)$ $p$-form symmetry. 
At the leading order of the functional-derivative expansion,  
the action takes the following form,  
\aln{\label{general action}
S[\psi]&={\cal N}\int [dC_p^{}]\left(-\frac{1}{\mathrm{Vol}[C_p^{}]}\int d^p\xi \sqrt{h}\frac{\delta \psi^\dag[C_p^{}]}{\delta X^\mu(\xi)} \frac{\delta \psi[C_p^{}]}{\delta X_\mu^{}(\xi)} 
-V(\psi^\dagger[C_p^{}]\psi[C_p^{}])\right)~,
}
where Vol$[C_p^{}]$ is the world volume~(\ref{world volume}) and ${\cal N}$ is a normalization factor determined later.  
As we mentioned in the previous section, the functional derivative contains various area derivatives of lower degrees in general. 
In this paper, we focus on the $p$-th order area derivative and consider the simplified version of Eq.~(\ref{general action}):
\aln{
\label{brane field action 1}
S_0^{}[\psi]&={\cal N}\int [dC_p^{}]\left(-\frac{1}{\mathrm{Vol}[C_p^{}]}\int_{\Sigma_d^{}} \delta(C_p^{})D\psi^\dagger[C_p] \wedge \star D\psi^{}[C_p]
-V(\psi^\dagger[C_p^{}]\psi[C_p^{}])\right)\notag
\\
&={\cal N}\int [dC_p^{}]\Biggl(-\frac{1}{(p+1)!\mathrm{Vol}[C_p^{}]}\int_{C_p^{}}d^{p}\xi\sqrt{h}  \frac{\delta \psi^\dagger[C_p^{}]}{\delta \sigma^{\mu_1^{}\cdots\mu_{p+1}^{}}(\xi)}\frac{\delta \psi[C_p^{}]}{\delta \sigma_{\mu_1^{}\cdots\mu_{p+1}^{}}(\xi)}\notag\\
&\qquad-V(\psi^\dagger[C_p^{}]\psi[C_p^{}])\Biggr)~,
}
where $\delta (C_p^{})$ is defined as
\aln{\label{eq:delta(Cp)}
\delta(C_p^{})\coloneqq\int_{C_p^{}}^{} d^p\xi\sqrt{-\frac{h}{g}}\prod_{\mu=0}^{d-1}\delta\left(X^\mu-X^\mu(\xi)\right)~.
}
Besides, the (path-)integral measure is defined by
\aln{
[dC_p^{}]={\cal D}X e^{-T_p^{}\mathrm{Vol}[C_p^{}]}~,
\label{measure}
}
where $T_p^{}$ is the $p$-brane tension, and ${\cal D}X$ is the induced measure by the diffeomorphism invariant norm~\cite{KAWAI199293},
\aln{||\delta X||^2\coloneqq\int d^p\xi\sqrt{h}g_{\mu\nu}^{}(X(\xi))\delta X^\mu(\xi)\delta X^\nu(\xi)~.
}
The weight in the (path-)integral measure (\ref{measure}) that we choose is nothing but the $p$-brane action~\cite{Leigh:1989jq}, which suppresses large branes.  
Besides, when $\{X^\mu(\xi)\}$ represents an embedding of a closed subspace $C_p^{}$, its translation $\{X^{\mu}(\xi)+X_0^\mu\}$ also represents another closed subspace, which means that there are always zero mode integrations in Eq.~(\ref{measure}).
Equation~\eqref{brane field action 1} is a straightforward generalization of the action for mean string-field theory~\cite{Iqbal:2021rkn} to $p$-dimensional brane.

The brane action~(\ref{brane field action 1}) is invariant under the global $\mathrm{U}(1)$ $p$-form transformation, 
\aln{
\psi[C_p^{}]~\rightarrow ~\exp\left(i\int_{C_p^{}}\Lambda_p^{}\right)\psi[C_p^{}]~,\quad d\Lambda_p^{}=0~.
\label{global p-form}
}
Note that if $C_p^{}$ is a boundary, i.e., $C_p=\partial C_{p+1}$, the contribution of $\Lambda_p$ to the phase vanishes from the Stokes theorem.
Even when the topology of space-time is trivial, this transformation is useful as a symmetry, e.g., it leads to Ward-Takahashi identities as in ordinary $p=0$ symmetry. 

We can also promote the global symmetry to the gauge symmetry by introducing a $(p+1)$-form gauge field,
\aln{B_{p+1}^{}=\frac{1}{(p+1)!}B_{p+1,\mu_1^{}\cdots \mu_{p+1}^{}}^{}(X)dX^{\mu_1^{}}\wedge \cdots \wedge dX^{\mu_{p+1}^{}}~
}
and replacing the derivative with the covariant derivative,
\aln{\label{covariant derivative}
\frac{D_G^{} \psi[C_p^{}]}{\delta \sigma^{\mu_1^{}\cdots \mu_{p+1}^{}}(\xi)}\coloneqq\left(\frac{\delta}{\delta \sigma^{\mu_1^{}\cdots \mu_{p+1}^{}}(\xi)}-iB_{p+1,\mu_1^{}\cdots \mu_{p+1}^{}}^{}(X(\xi))\right)\psi[C_p^{}]~.
}
The gauge transformation is given by
\aln{
\psi[C_p^{}]~\rightarrow~e^{i\int_{C_p^{}}\Lambda_p^{}}\psi[C_p^{}]~,\quad B_{p+1}^{}~\rightarrow~B_{p+1}^{}+d\Lambda_p^{}~,
}
where we have used Eq.~(\ref{Wilson loop are derivative}). 
Note that the action~(\ref{brane field action 1}) is invariant under the spacetime diffeomorphism and the reparametrization on $C_p^{}$ by the construction of $\psi[C_p^{}]$.  

\

We should comment on the interactions of the brane field. 
The potential $V(\psi^\dagger[C_p^{}]\psi[C_p^{}])$ in Eq.~(\ref{brane field action 1}) corresponds to contact interactions in ordinary field theory, but we can also consider more general interactions such as~\cite{Iqbal:2021rkn} 
\aln{\label{topology changing interaction}
\int [dC_p^1]\int [dC_p^2]\int [dC_p^3]\delta(C_p^1-C_p^2-C_p^3)\psi^\dagger[C_p^1]\psi[C_p^2]\psi[C_p^3]+{\rm h.c.}~,
}
which represents the splitting or merging of branes\footnote{This interaction still preserves the $\mathrm{U}(1)$ $p$-form symmetry due to the delta function. 
It may seem strange to have $\mathrm{U}(1)$ symmetry while the number of branes is changing; however, $\mathrm{U}(1)$  $p$-form symmetry does not represent the conservation of the number of branes itself, but rather the conservation of the winding number on a space with nontrivial topology.
}.
Such interactions seem to alter the mean-field dynamics of the brane field significantly as the behaviors of phase transition in the ordinary Landau theory change by adding odd potential terms. 
In this paper, we simply neglect these interactions and focus on the model~(\ref{brane field action 1}).

\subsection{Conservation law}\label{sec:Conservation law}
As with ordinary symmetry,
when $p$-form global symmetry is continuous, we have a current $(p+1)$-form $J_{p+1}^{}$, which is conserved as 
\aln{d\star J_{p+1}^{}=0~.
}
The corresponding conserved charge is given as
\aln{
Q_p^{}[\Sigma_{d-p-1}]=\int_{\Sigma_{d-p-1}^{}} \star J_{p+1}^{}~,
}
where $\Sigma_{d-p-1}^{}$ is a $(d-p-1)$-dimensional (closed) subspace.   

We can calculate $ J_{p+1}^{}$ in the brane field theory as follows.  
Instead of the global $p$-form transformation, we consider an infinitesimal local $p$-form transformation
\aln{
\psi[C_p^{}]\quad \rightarrow \quad e^{i\int_{C_p^{}}\Lambda_p^{}}\psi[C_p^{}]~,\quad d\Lambda_p^{}\neq 0~.
\label{eq:field transform}
}
Then, the variation of the action in the linear order of $\Lambda_p^{}$ should have the form
\begin{equation}
    \delta S =-\int_{\Sigma_d}d\Lambda_p\wedge \star J_{p+1}~,
    \label{eq:Noether}
\end{equation}
since the action needs to vanish if $d\Lambda_p=0$.
$J_{p+1}$ is nothing but the Noether current.
From the integrating by part, we obtain
\begin{equation}
    \delta S =(-)^{p}\int_{\Sigma_d}\Lambda_p\wedge d\star J_{p+1}~.
    \label{eq:delta_S_current}
\end{equation}
If $\psi$ satisfies the equation of motion, the action is stationary, so the divergence of the current vanishes $d\star J_{p+1}=0$.  

Let us derive the explicit form of $J_{p+1}$ for the action~(\ref{brane field action 1}).
The variation of the action is calculated as
\aln{
\delta S_0^{}[\psi]&={\cal N}\int [dC_p^{}]\frac{
i}{\mathrm{Vol}[C_p^{}]}\int_{\Sigma_d^{}} \delta(C_p^{})\left(d\Lambda_p^{}\wedge \star \psi^\dagger D\psi-\psi D\psi^\dagger\wedge \star d\Lambda_p^{}\right)
\notag\\
&={\cal N}\int [dC_p^{}]\frac{
i}{\mathrm{Vol}[C_p^{}]}\int_{\Sigma_d^{}} \delta(C_p^{})d\Lambda_p^{}\wedge \star (\psi^\dagger D\psi-\psi D\psi^\dagger)~,
\label{eq:delta_S}
} 
where we have used $\eta\wedge \star \omega=\omega \wedge \star \eta$. 
Comparing Eqs.~\eqref{eq:Noether} and \eqref{eq:delta_S}, we obtain
\aln{
J_{p+1}^{}(X)=-
{\cal N}\int [dC_p^{}]\frac{\delta(C_p^{})}{\mathrm{Vol}[C_p^{}]}i(\psi^\dagger D\psi-\psi D\psi^\dagger)~.
\label{p+1 current}
} 
This expression shows that the $(p+1)$-form current is given by the integral over all the brane configurations. 
Note that $X$ dependence of the current $J_{p+1}(X)$ comes from $\delta(C_p)$ in Eq.~\eqref{eq:delta(Cp)}.
We can also check that $Q_p^{}$ generates the $p$-form transformation~(\ref{global p-form}) in the following way.  
We formally define the quantum theory of the present brane field by the following path integral: 
\aln{
\langle {\cal O}(X(\xi))\rangle\coloneqq\frac{1}{Z}\int {\cal D}\psi {\cal D}\psi^\dagger  {\cal O}(X(\xi))e^{iS^{}[\psi]}~. 
}
Consider the expectation value of $\psi[C_p]$,
\begin{equation}
\langle \psi[C_p] \dots\rangle~,
\end{equation}
where ``$\dots$'' represents other arbitrary operators.
We choose $\Lambda_p$
in the field transformation of Eq.~\eqref{eq:field transform} as 
\begin{equation}
    \Lambda_p = \epsilon (-1)^{p(d-p-1)}\delta_p(D_{d-p})
    \label{eq:Lambda_p}
\end{equation}
with $\partial D_{d-p}=\Sigma_{d-p-1}$, and an infinitesimal parameter $\epsilon$.
Here, $\delta_p(D_{d-p})$ is the Poincar\'e-dual form of $D_{d-p}$ such that
\begin{equation}
\int_{D_{d-p}} f_{d-p} = \int_{\Sigma_d} f_{d-p} \wedge \delta_{p}^{}(D_{d-p}^{})~,
\label{eq:Poincare dual form}
\end{equation}
for an arbitrary $(d-p)$-form, $f_{d-p}$. 
$\delta_{p}^{}(D_{d-p}^{})$ can be thought of as a generalization of the delta function.

\begin{figure}
    \centering
    \includegraphics[scale=0.4]{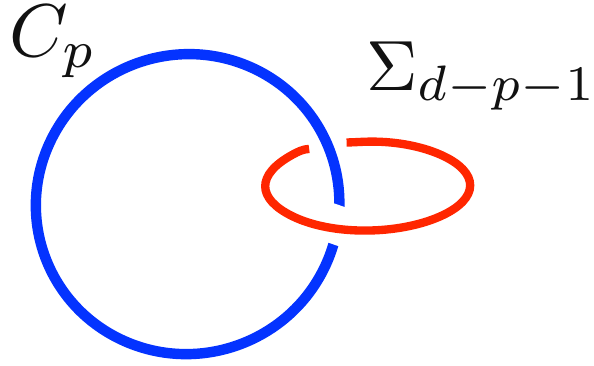}
    \qquad\qquad\includegraphics[scale=0.6]{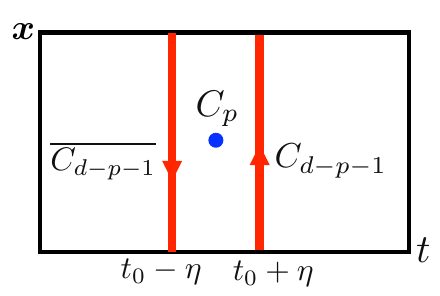}
    \caption{(Left) Configuration of the link between $C_p$ and $\Sigma_{d-p-1}$ for $d=3$ and $p=1$.
    (Right) Configuration of $C_p$ and $C_{d-p-1}$ to obtain the commutation relation in operator formalism.
    The arrows indicate the orientation of the subspace.}
    \label{fig:link}
\end{figure}

In this parametrization, the variance of the action~\eqref{eq:Noether} is
\begin{equation}
    \delta S =  
    \epsilon \int_{\Sigma_{d-p-1}}\star J=
    \epsilon Q_p[\Sigma_{d-p-1}]~,
\end{equation}
while $\psi[C_p^{}]$ transforms as 
\begin{equation}
    \psi[C_p^{}] \rightarrow  e^{i\int_{C_p^{}}\Lambda_p^{}}\psi[C_p^{}]
    =e^{i\epsilon (-1)^{p}\mathrm{Link}[\Sigma_{d-p-1}^{},C_p^{}]}\psi[C_p^{}]
    ~.
\end{equation}
Here, we have defined the linking number as 
\begin{equation}
\mathrm{Link}[\Sigma_{d-p-1}^{},C_p^{}]\coloneqq\int_{\Sigma_d} \delta_{d-p}(C_p^{})\wedge \delta_{p}(D_{d-p})~.
\label{eq:Linking number}
\end{equation}
See the left figure in Fig.~\ref{fig:link} for an example of the configuration of links, where we take $d=3$ and $p=1$.
We assume other operators in ``$\dots$'' do not have support on $D_{d-p}$, so that ``$\dots$'' does not transform under the field transformation with Eq.~\eqref{eq:Lambda_p}.
In the path-integral, the field transformation~\eqref{eq:field transform} is merely a redefinition of the integral variables. Therefore, assuming the path-integral measure is invariant under the field transformation~\eqref{eq:field transform}, it leads to the identity,
\begin{equation}
    \langle \psi[C_p] \dots\rangle = 
    \langle e^{i\epsilon Q_p[\Sigma_{d-p-1}] + i\epsilon (-1)^{p}\mathrm{Link}[\Sigma_{d-p-1}^{},C_p^{}]} \psi[C_p] \dots\rangle~.
\end{equation}
This implies
\begin{equation}
    Q_p[\Sigma_{d-p-1}]\psi[C_p] = -(-1)^{p}\mathrm{Link}[\Sigma_{d-p-1}^{},C_p^{}]\psi[C_p]~.
    \label{eq:path integral relation}
\end{equation}
This is the relation in the path integral.
To derive the relation in the operator formalism, consider the Cauchy surface labeled by time $t$. 
Let $C_{d-p-1}(t_0)$ be $(d-p-1)$-dimensional subspace on the Cauchy surface at $t=t_0$.
We choose that $\Sigma_{d-p-1}=C_{d-p-1}(t_0+\eta)\cup\overline{C_{d-p-1}}(t_0-\eta)$ with an infinitesimal parameter $\eta$.
Here, $\overline{C_{d-p-1}}$ is the $(d-p-1)$-dimensional subspace with the opposite orientation of $C_{d-p-1}$.
We also choose that $C_p$ is a $p$-dimensional subspace on the Cauchy surface at $t=t_0$.
See the right figure in Fig.~\ref{fig:link} for the configuration.
In this configuration, the left-hand side in Eq.~\eqref{eq:path integral relation} becomes
\begin{equation}
    Q_p[\Sigma_{d-p-1}]\psi[C_p]=Q_p[C_{d-p-1}(t_0+\eta)]\psi[C_p]-Q_p[C_{d-p-1}(t_0-\eta)]\psi[C_p]~.
\end{equation}
In operator formalism, the ordering of the operator product corresponds to the time ordering, so 
$Q_p[C_{d-p-1}(t_0+\eta)]\psi[C_p]-Q_p[C_{d-p-1}(t_0-\eta)]\psi[C_p]\to [Q_p[C_{d-p-1}],\psi[C_p] ]$; thus, we find
that the Noether charge $Q_p$ generates the symmetry transformation,
\begin{equation}
        [i Q_p[C_{d-p-1}],\psi[C_p] ]= -i \mathrm{I}[C_{d-p-1}^{},C_p^{}]\psi[C_p]~.
        \label{eq:operator relation2}
\end{equation}
Here, $\mathrm{I}[C_{d-p-1}^{},C_p^{}]$ represents the intersection number between $C_{d-p-1}^{}$ and $C_p^{}$ on the Cauchy surface, which can be obtained by evaluating the right-hand side in Eq.~\eqref{eq:path integral relation} with $\Sigma_{d-p-1}=C_{d-p-1} (t_0+\eta)\cup\overline{C_{d-p-1}}(t_0-\eta)$.
%
%

\section{Spontaneous breaking of Higher-form symmetry}\label{sec:SSB}
In this section, we discuss the spontaneous breaking of the higher-form symmetry in the brane field theory. 
As in the case of $0$-form symmetry, gapless modes appear when the continuous $p$-form symmetry is spontaneously broken~\cite{Gaiotto:2014kfa,Lake:2018dqm,Hofman:2018lfz,Hidaka:2020ucc}.
For $0$-form symmetry, the symmetry breaking is characterized by an order parameter that is 
the expectation value of a local field $\langle\phi(x)\rangle$. 
This order parameter cannot be directly extended to higher-form symmetries.
Alternatively, we can use the off-diagonal long-range order,
\begin{equation}
O_0=\lim_{|x-y|\to\infty}\langle \phi^\dag(x)\phi(y)\rangle
\simeq \langle \phi^\dag(x)\rangle\langle\phi(y)\rangle~,
\end{equation}
as the order parameter. Since the two points $(x,y)$ can be written as the boundary of segment $M_1$ and the distance between $x$ and $y$ can be expressed as  the minimal volume $|x-y|=\mathrm{Vol}(M_1)$, 
the order parameter can be written in the form:
\begin{equation}
    O_0 = \lim_{\mathrm{Vol}[M_1]\to\infty}\langle \phi^\dag(x)\phi(y)\rangle~.
\end{equation}
This expression can be naturally extended to accommodate the case of $p$-form symmetry. 
In this context, we can define the order parameter as
\begin{equation}
O_p=\lim_{\mathrm{Vol}[M_{p+1}]\to\infty} \langle \psi[C_p]\rangle~,
\label{eq:Op}
\end{equation}
where $\partial M_{p+1}=C_p$.
We will use Eq.~\eqref{eq:Op} as the order parameter of spontaneous breaking of $p$-form symmetry. 
In general, the order parameter defined in Eq.~\eqref{eq:Op} might vanish in the limit of large $M_{p+1}$, depending on $C_p=\partial M_{p+1}$, i.e., the perimeter law. 
In such cases, it is necessary to consider a renormalized order operator $\langle Z(C_p)\psi[C_p]\rangle$ with a field-independent functional $Z(C_p)$. If the order operator does not vanish no matter what renormalization is performed, we can say that the symmetry is spontaneously broken.

We work within the mean-field approximation. 
As an ansatz for the solution, we assume that the brane-field configuration $\psi[C_p]$ depends only on the minimum volume with the boundary $C_p$,
\aln{
\psi[C_p^{}]=\frac{1}{\sqrt{2}}f(z=\mathrm{Vol}[M_{p+1}^{}])~,\quad \partial M_{p+1}^{}=C_p^{}~.
}
This corresponds to the truncated treatment in Ref.~\cite{Iqbal:2021rkn}.  
See also Appendix~\ref{sec:truncation} for more general truncations. 
By using Eq.~(\ref{area derivative of volume}), 
the area derivative becomes 
\aln{D\psi[C_p^{}]=\frac{1}{\sqrt{2}}E_{p+1}^{}f'(z)~,
}
which leads to
\aln{
D\psi^\dagger \wedge \star D\psi
=\frac{1}{2}f'(z)^2 E_p\wedge \star E_p
=\sqrt{-g}\frac{1}{2}f'(z)^2
dX^{0}\wedge\cdots\wedge dX^{d-1}
~.
}
Here, we used Eq.~\eqref{EE} to evaluate $E_p\wedge \star E_p$.
Then, the action~\eqref{brane field action 1} becomes
\aln{
S_0^{}[f]&=-{\cal N}\int [dC_p^{}]\left(\frac{f'(z)^2}{2\mathrm{Vol}[C_p^{}]}\int_{C_p^{}}d^p\xi\sqrt{h} +V(f(z)^2) 
\right)\bigg|_{z=\mathrm{Vol}[M_{p+1}^{}]}^{}
\notag\\
&=-{\cal N}\int {\cal D}X e^{-T_p^{}\mathrm{Vol}[C_p^{}]}
\left(\frac{1}{2}f'(z)^2+V(f(z)^2) 
\right)\bigg|_{z=\mathrm{Vol}[M_{p+1}^{}]}^{}
\notag\\
&=-\int_0^\infty dzg(z)\left(\frac{1}{2}f'(z)^2+V(f(z)^2) 
\right)~,
}
where 
\aln{
g(z)={\cal N}\int {\cal D}X e^{-T_p^{}\mathrm{Vol}[C_p^{}]}\delta(z-\mathrm{Vol}[M_{p+1}^{}])
}
is the density of $p$-brane configurations for a given minimal volume $z$.  
The equation of motion for $f(z)$ is 
\aln{
&\frac{d}{dz}\left(g(z)f'(z)\right)-g(z)\frac{\partial V}{\partial f}=0~,
\\
\therefore\quad & f''(z)+\frac{g'(z)}{g(z)}f'(z)-\frac{\partial V}{\partial f}=0~.
\label{truncated eom}
}  
By introducing the WKB form $f(z)=\exp(S(z))$, Eq.~(\ref{truncated eom}) can be also rewritten as 
\aln{
S''(z)+S'(z)^2+\frac{g'(z)}{g(z)}S'(z)-\frac{1}{f(z)}\frac{\partial V}{\partial f}=0~.
} 
As in the usual Landau theory, the vacuum state is determined by the potential $V(f^2)$. 
As a simplest example, let us consider the following potential:
\aln{V(f^2)&=m(\psi^\dagger \psi)+\frac{\lambda}{4}(\psi^\dagger \psi)^2\notag
\\
&=\frac{m}{2}f^2+\frac{\lambda}{8}f^4~.
}
In the following, we always assume $\lambda>0$ to guarantee the stability of the system.

\subsection{Unbroken phase}\label{sec:unbroken}
When $m>0$, the minimum of the potential is located at $f=0$.
Therefore, we can neglect the quartic potential in the equation of motion as
\aln{S''(z)+S'(z)^2+\frac{g'(z)}{g(z)}S'(z)-m\approx 0~.
}
Let us find the asymptotic solution for the large volume.
For $z\rightarrow \infty$, we have $g'(z)/g(z)\sim z^{-1}$ from dimensional analysis\footnote{Actually, this dimensional analysis is not fully correct because there also exists another dimensionful quantity, i.e., the brane tension $T_p^{}$. 
Roughly speaking, the large $z$ limit would behave as $g(z)\sim  \exp\left(-T_p^{}z^{\frac{p}{p+1}}\right)$, which means $g'/g\sim z^{-\frac{1}{p+1}}$. 
In any case, the large $z$ behavior of $f(z)$ does not change. 
} and the solution is given by
\aln{
S[z]\approx-\sqrt{m}z~,
}
which corresponds to the area law of the brane field  
\aln{\label{area low}
\psi[C_p^{}]=c\exp\left(-\sqrt{m}\mathrm{Vol}[M_{p+1}^{}]\right)\quad \text{for Vol$[M_{p+1}^{}]\rightarrow \infty$}~,
}
where $c$ is a constant that in principle can be determined if we specify the boundary condition for $z\rightarrow 0$ (small brane limit). 
The exponential behavior also justifies neglecting the quartic potential term in the equation of motion. 
This implies that the order parameter vanishes, indicating the unbroken phase of $p$-form symmetry.

Equation~(\ref{area low}) should be compared to the correlation function of the ordinary field theory ($p=0$):
\aln{
\langle \phi^\dag(x)\phi(y)\rangle\sim e^{-\frac{|x-y|}{\xi(m)}}~,\quad \xi(m)\propto \frac{1}{m^{1/2}}~.
}
In the $p$-form case, Eq.~(\ref{area low}) means 
\aln{
(\text{volume tension})\sim m^{1/2}
}
within the mean-field approximation. 
 
\subsection{Broken phase}\label{sec:broken}

Let us next consider the case for $m<0$.
Within the truncated approximation, the equation of motion is given by Eq.~(\ref{truncated eom}): 
\aln{
f''(z)+\frac{g'(z)}{g(z)}f'(z)-mf(z)-\frac{\lambda}{2}f(z)^3=0~.
}
As in the unbroken case, we focus on the large $z$ behavior. In this case, we can neglect the derivative terms for $z\rightarrow \infty$ by dimensional analysis, and the solution is given by $f(z)^2=v^2\coloneqq-2m/\lambda$. 
Since the order parameter $\psi[C_p^{}]=v/\sqrt{2}$ is nonvanishing at $z\to\infty$, the $p$-form symmetry is spontaneously broken. 

In ordinary quantum field theory, the non-renormalized order parameter exhibits a perimeter law in the broken phase.
However, the order parameter is completely independent of $C_p^{}$ in the present model. 
As already mentioned in Ref.~\cite{Iqbal:2021rkn},  this may be an artifact by having neglected the topology-changing terms such as Eq.~(\ref{topology changing interaction}). 
It would be interesting to see whether we can actually realize the perimeter law by adding such topology-changing interactions. 
 
\subsection{Nambu-Goldstone modes}\label{sec:NG mode}
What are the low-energy fluctuation modes in the broken phase? 
%
As in the case of $0$-form symmetry, the phase fields are candidates for low-energy degrees of freedom (d.o.f):
\aln{\psi[C_p^{}]=\frac{v}{\sqrt{2}}\exp\left(i\int_{C_p^{}}A_p^{}\right)~,\quad A_p^{}=\frac{1}{p!}A_{p,\mu_1^{}\cdots \mu_p^{}}^{}(X)dX^{\mu_1^{}}\wedge \cdots \wedge dX^{\mu_p^{}}~.
\label{NG mode}
}
Let us see how this d.o.f describes a gapless mode in the effective action.
Note that the effective theory has the gauge symmetry 
\aln{A_p^{}\quad \rightarrow \quad A_p^{}+d\Lambda_{p-1}^{}~,\quad d\Lambda_{p-1}^{}\neq 0~,
}
because Eq.~(\ref{NG mode}) is invariant under this transformation due to the closedness of $C_p^{}$.
Here $\Lambda_{p-1}$ is $(p-1)$-form gauge parameter. 
For Eq.~\eqref{NG mode} to be invariant, the integral of $d\Lambda_{p-1}$ need not vanish, but can be
\begin{equation}
    \int_{C_p} d\Lambda_{p-1} \in 2\pi \mathbb{Z}~.
\end{equation}
In other words, $A_p$ is the $\mathrm{U}(1)$ $p$-form gauge field.

Now let us calculate the effective action for $A_p^{}$. 
By putting Eq.~(\ref{NG mode}) into the action~(\ref{brane field action 1}) and using Eq.~(\ref{Wilson loop are derivative}), we have
\aln{
S_0^{}[A_p^{}]=-{\cal N}\int [dC_p^{}]\left(\frac{v^2}{2(p+1)!\mathrm{Vol}[C_p^{}]}\int_{C_p^{}}d^{p}\xi \sqrt{h} F^{\mu_1^{}\cdots \mu_{p+1}^{}}(X(\xi))F^{}_{\mu_1^{}\cdots \mu_{p+1}^{}}(X(\xi))
\right)~.
\label{effective action 1}
}
In the following, we consider the flat spacetime $g_{\mu\nu}^{}=\eta_{\mu\nu}^{}$ for simplicity. 
By introducing the Fourier modes, 
\aln{F(x)\coloneqq v^2F^{\mu_1^{}\cdots \mu_{p+1}^{}}(x)F^{}_{\mu_1^{}\cdots \mu_{p+1}^{}}(x)=\int \frac{d^dk}{(2\pi)^d}e^{ik\cdot x}\tilde{F}(k)~,
}
Eq.~(\ref{effective action 1}) can be written as
\aln{
S_0^{}[A_p^{}]=-\int \frac{d^dk}{(2\pi)^d}K(k)\tilde{F}(k)~,
}
where
\aln{K(k)={\cal N}\int {\cal D}X\frac{1}{2(p+1)!\mathrm{Vol}[C_p^{}]}\int_{C_p^{}}d^{p}\xi \sqrt{h} e^{-T_p^{}\mathrm{Vol}[C_p^{}]+ik_\mu^{}X^\mu(\xi)}~.
}
There are zero-mode integrations in the above path-integral, and it is convenient to separate them:
\aln{K(k)={\cal N}\int d^dx e^{ik\cdot x}\int {\cal D}X_\mathrm{NZ}\frac{1}{2(p+1)!\mathrm{Vol}[C_p^{}]}\int_{C_p^{}}d^{p}\xi\sqrt{h} e^{-T_p^{}\mathrm{Vol}[C_p^{}]+ik_\mu^{}X_\mathrm{NZ}^\mu(\xi)}~,
}
where $X_\mathrm{NZ}^\mu(\xi)$ denotes the nonzero mode. 
In this expression, it is easy to see that $K(k)$ is proportional to $\delta^{(d)}(k)$.  
Thus, we can choose the normalization ${\cal N}$ as
\aln{
{\cal N}\int {\cal D}X_\mathrm{NZ}\frac{1}{\mathrm{Vol}[C_p^{}]}\int_{C_p^{}}d^{p}\xi\sqrt{h} e^{-T_p^{}\mathrm{Vol}[C_p^{}]}=1~\leftrightarrow~K(k)=\frac{(2\pi)^d}{2(p+1)!}\delta^{(d)}(k)~,
\label{normalization}
} 
which leads to 
\aln{\label{eq:S0}
S_0^{}[A_p^{}]&=-\frac{1}{2(p+1)!}\tilde{F}(0)\notag\\
&=-\frac{v^2}{2(p+1)!}\int d^dx F^{\mu_1^{}\cdots \mu_{p+1}^{}}(x)F^{}_{\mu_1^{}\cdots \mu_{p+1}^{}}(x)\notag\\
&=-\frac{v^2}{2}\int_{\Sigma_d^{}} F_{p+1}^{}\wedge \star F_{p+1}^{}~,
}
which is nothing but the $p$-form Maxwell theory, and thus, $A_p^{}$ is gapless for $d>p+2$\footnote{For $d\leq p+2$, $\mathrm{U}(1)$ $p$-form symmetry cannot be broken by the higher-form version of the Mermin-Wagner theorem~\cite{Gaiotto:2014kfa,Lake:2018dqm}.
}
. 
%

\subsection{Other fluctuation modes}
What about the other fluctuation modes?
In general, they are given by expanding the phase with respect to the derivatives of $X^\mu(\xi)$:
\aln{
\psi[C_p^{}]&=\frac{v}{\sqrt{2}}\exp\left(i\int_{C_p^{}}d^p\xi\sqrt{h}\left\{\phi(X(\xi))+h^{ij}(\xi)\frac{\partial X^\mu}{\partial\xi^i}\frac{\partial X^\nu}{\partial\xi^j}H_{\mu\nu}^{}(X(\xi))+\cdots\right\}\right)~,
}
where $\phi(X)$ is a scalar field and $H_{\mu\nu}^{}(X)$ is a symmetric tensor field. 
These fluctuation modes are typically gapped because they are not protected by spontaneous symmetry breaking. 

Here, we actually show that $\phi(X)$ is gapped as an example. 
The area derivative is calculated as 
\aln{D\int_{C_p^{}}d^p\xi \sqrt{h}\phi(X(\xi))=d(\phi E_{p}^{})=\frac{(p+1)}{(p+1)!}\partial_{[\mu_1^{}}^{}\left(\phi E_{\mu_2^{}\cdots \mu_{p+1}^{}]}^{}\right)dX^{\mu_1^{}}\wedge \cdots \wedge dX^{\mu_{p+1}^{}}~.
}
Then, we have 
\aln{
D\psi^\dagger \wedge \star D\psi=\frac{v^2\sqrt{-g}(p+1)^2}{2(p+1)!}\partial_{[\mu_1^{}}^{}\left(\phi E_{\mu_2^{}\cdots \mu_{p+1}^{}]}^{}\right)\partial^{[\mu_1^{}}\left(\phi E^{\mu_2^{}\cdots \mu_{p+1}^{}]}\right)dX^0\wedge \cdots dX^{d-1}~,
} 
Now, the effective action is written as 
\aln{
S_0^{}[\phi]&=-{\cal N}\int [dC_p^{}]\frac{v^2}{2\mathrm{Vol}[C_p^{}]}\int_{C_p^{}}d^p\xi \sqrt{h}\notag\\
&\quad\times\left(\frac{(p+1)^2}{(p+1)!}(\partial_{[\mu_1^{}}^{}\phi) E_{\mu_2^{}\cdots \mu_{p+1}^{}]}( \partial^{[\mu_1^{}}\phi)E^{\mu_2^{}\cdots \mu_{p+1}^{}]}
+M^2\phi^2+2\phi \partial_\mu^{}\phi G^\mu
\right)~,\label{effective action of scalar}
}
where 
\aln{
G^\mu(X(\xi))&=\frac{(p+1)^2}{(p+1)!}E_{\mu_1^{}\cdots \mu_{p}^{}}^{}\partial^{[\mu^{}}E^{\mu_1^{}\cdots \mu_{p}^{}]}~,\\
 M^2(X(\xi))&=\frac{(p+1)^2}{(p+1)!}\partial_{[\mu_1^{}}^{}E_{\mu_2^{}\cdots \mu_{p+1}^{}]}^{}\partial^{[\mu_1^{}}E^{\mu_2^{}\cdots \mu_{p+1}^{}]}~.
}
The first term in Eq.~\eqref{effective action of scalar} can be expressed as 
\aln{
&\frac{(p+1)^2}{(p+1)!}(\partial_{[\mu_1^{}}^{}\phi) E_{\mu_2^{}\cdots \mu_{p+1}^{}]}( \partial^{[\mu_1^{}}\phi)E^{\mu_2^{}\cdots \mu_{p+1}^{}]}\notag\\
&=\frac{1}{(p+1)!}\Bigl(
(\partial_{\mu_1^{}}^{}\phi) E_{\mu_2^{}\mu_3\cdots \mu_{p+1}^{}}
-(\partial_{\mu_2^{}}^{}\phi) E_{\mu_1^{}\mu_3\cdots \mu_{p+1}^{}}
\cdots 
-(\partial_{\mu_{p+1}^{}}^{}\phi) E_{\mu_2^{}\mu_3\cdots \mu_{1}^{}}
\Bigr)\notag\\
&\quad\times
\Bigl(
(\partial^{\mu_1^{}}\phi) E^{\mu_2^{}\mu_3\cdots \mu_{p+1}^{}}
-(\partial^{\mu_2^{}}\phi) E^{\mu_1^{}\mu_3\cdots \mu_{p+1}^{}}
\cdots 
-(\partial^{\mu_{p+1}}\phi) E^{\mu_2^{}\mu_3\cdots \mu_{1}^{}}
\Bigr)\notag\\
&=\frac{1}{(p+1)!}\Bigl(
(p+1)(\partial_{\mu^{}}^{}\phi)(\partial^{\mu^{}}\phi)  E_{\mu_1^{}\mu_2\cdots \mu_{p}^{}}E^{\mu_1^{}\mu_2\cdots \mu_{p}^{}}\notag\\
&\qquad\qquad\qquad-p(p+1)(\partial_{\mu^{}}^{}\phi)(\partial^{\nu^{}}\phi) E_{\nu^{}\mu_1\mu_2\cdots \mu_{p-1}^{}}E^{\mu^{}\mu_1\mu_2\cdots \mu_{p-1}^{}}
\Bigr)\notag\\
&=
(g^{\mu\nu}-e_i^{\mu}e_j^{\nu}h^{ij})(\partial_{\mu^{}}^{}\phi)(\partial_{\nu^{}}^{}\phi)
\notag\\
&=n^\mu n^\nu(\partial_{\mu^{}}^{}\phi)(\partial_{\nu^{}}^{}\phi)~,
\label{scalar kinetic term}
}
where $n^\mu$ is the normal vector on $C_p^{}$, and we have used $g^{\mu\nu}=n^\mu n^\nu+h^{ij}e_i^\mu e_j^\nu$ in the last line.  
Now let us focus on the flat spacetime $g_{\mu\nu}^{}=\eta_{\mu\nu}^{}$ for simplicity.  
Equation~(\ref{scalar kinetic term}) gives the kinetic term 
\aln{\label{effective kinetic term}
-\frac{1}{2}\int d^dx \langle n^\mu n^\nu\rangle \partial_\mu^{}\phi \partial_\mu^{}\phi~, 
}
where 
\aln{\label{mixing term}
\langle n^\mu n^\nu\rangle={\cal N}\int {\cal D}X_{\rm NZ}^{}\frac{v^2}{\mathrm{Vol}[C_p^{}]}\int_{C_p^{}}d^{p}\xi \sqrt{h}e^{-T_p^{}\mathrm{Vol}[C_p^{}]}n^\mu(\xi)n^\nu(\xi)~,
}
which has to be proportional to $\eta^{\mu\nu}$ as long as the Lorentz symmetry is unbroken.\footnote{
A rigorous proof needs more dedicated studies. 
Instead, we here give an intuitive argument. 
For a given $C_p^{}$ and $\mu\neq \nu$, there always exists another brane $C_p^{'}$ such that $n'^\mu=-n^\mu$ and $n'^\nu=n^\nu$ for $\nu\neq \mu$. 
These contributions cancel each other in the path-integral (\ref{mixing term}), and it should vanish for $\mu\neq \nu$.   
We can show this more explicitly based on the gauge-fixed brane action for  $p=0,1$~\cite{Iqbal:2021rkn}.  
} 
As a result, Eq.~(\ref{effective kinetic term}) gives the usual kinetic.  

For the evaluation of the other terms in Eq.~(\ref{effective action of scalar}), note that $E_p^{}$ (and correspondingly $G^\mu$ and $M^2$) does not depend on the spacetime zero mode $x^\mu$ by definition. 
Then, the second term becomes the mass term
\aln{
-\frac{v^2\Lambda^2}{2}\int d^dx \phi(x)^2~,
} 
where $\Lambda^2$ is defined by
\aln{
\Lambda^2={\cal N}\int {\cal D}X_\mathrm{NZ}\frac{1}{\mathrm{Vol}[C_p^{}]}\int_{C_p^{}}d^{p}\xi \sqrt{h}e^{-T_p^{}\mathrm{Vol}[C_p^{}]}M^2(X_\mathrm{NZ}^{}(\xi))~,
}
which can be interpreted as a bare mass of $\phi(x)$. 
As for the third term, it contains the following average: 
\aln{
\langle G^\mu\rangle={\cal N}\int {\cal D}X_\mathrm{NZ}\frac{1}{\mathrm{Vol}[C_p^{}]}\int_{C_p^{}}d^{p}\xi \sqrt{h}e^{-T_p^{}\mathrm{Vol}[C_p^{}]}G^\mu(X_\mathrm{NZ}^{}(\xi))~. 
}  
This term should vanish as long as the spacetime symmetry is not spontaneously broken.  
Now, one can see that the scalar mode is gapped. 
Other fluctuation modes also become gapped as long as they are not protected by some additional symmetry of the original brane-field theory. 
%

\subsection{Emergent higher-form symmetry
}
In the case of ordinary $0$-form symmetries, there exists a vortex solution in the broken phase, which carries the topological charge given by
\aln{
Q_{d-2}^{}=\frac{1}{2\pi v^{2}}\oint_{S_1^{}} J_1^{}~,\quad J_1^{}=j_\mu^{}dx^\mu~,\quad j_\mu^{}=-i(\phi^*\partial_\mu^{}\phi-\phi\partial_\mu^{}\phi^*)~,
}
where $\phi$ is a complex scalar field, and $S_1^{}$ is a closed curve. 
This symmetry is not exact but rather emergent, as this charge is not strictly conserved,
\begin{equation}
    d J_1= -2id\phi^*\wedge d\phi~.
\end{equation}
If we parametrize $\phi=he^{i\varphi}/\sqrt{2}$, where $h$ and $\varphi$ are the radial and phase degrees of freedom, respectively, the current is expressed as $J_1=h^2 d\varphi$, leading to $d J_1=  2h dh\wedge d\varphi$. 
Away from the vortex core, $h$ approaches the vacuum expectation value (VEV) $h=v$, and the fluctuation of $h$ can be neglected at low energy because it is gapped. Consequently, $J_1$ can be approximated as $J_1\approx v^2d\varphi$, and the topological charge reduces to 
\aln{Q_{d-2}^{}=\oint \frac{d\varphi}{2\pi }~,
}
which is conserved due to $dd\varphi=0$.
$U(\theta)=e^{i\theta Q_{d-2}^{}}$ is the corresponding symmetry operator, which acts on  $(d-2)$-dimensional object,
i.e., $\mathrm{U}(1)$ $(d-2)$-form symmetry.
For example, for $d=4$, this is $2$-form symmetry, and the $2$-dimensional object is the worldsurface of a vortex.

In the case of $p$-form symmetry, a natural generalization of topological charge is given by 
\aln{
Q_{d-p-2}^{}=\frac{1}{2\pi v^{2}}\int_{S_{p+1}^{}}J_{p+1}^{}~,  
\label{winding number}
}
where $S_{p+1}^{}$ is a closed $(p+1)$-dimensional subspace and $J_{p+1}^{}$ is given in Eq.~\eqref{p+1 current}. 
%
By substituting Eq.~\eqref{NG mode} into 
Eq.~\eqref{p+1 current}, we obtain 
$J_{p+1}=v^2 F_{p+1}$, which can also be derived from the low-energy effective action~\eqref{eq:S0}, using the Noether theorem.
Consequently, the topological charge~\eqref{winding number} becomes
\aln{
Q_{d-p-2}^{}=\int_{S_{p+1}^{}}\frac{{F}_{p+1}^{}}{2\pi}~.
\label{topological Q}
}
%
The corresponding symmetry operator is $U(\theta)=e^{i\theta Q_{d-p-2}^{}}$ and the charged object is a $(d-p-2)$-dimensional object.
%
For example, when $d=4$ and $p=1$, the charged object is a worldline of a magnetic particle.  
%
Since $A_p$ is the $\mathrm{U}(1)$ $p$-form gauge field, it satisfies the Dirac quantization condition,
\begin{equation}
    \int_{S_{p+1}} F_{p+1} \in 2\pi\mathbb{Z}~,
    \label{eq:1st Chern class}
\end{equation}
which also leads to
\aln{
Q_{d-p-2}^{} \in \mathbb{Z}~.
}

\subsection{Discrete higher-form symmetry breaking
}\label{sec:Discrete}

Up to this point, we have discussed the case with a continuous higher-form symmetry. 
In general, we can consider a model with a discrete higher-form symmetry and its spontaneous breaking.
For example, we can construct a model with $\mathbb{Z}_N^{}$ $p$-form symmetry,
by adding the following term
\aln{
-\lambda_N^{}\left\{(\psi[C_p^{}])^N+(\psi^\dagger [C_p^{}])^N\right\}~,
\label{breaking term}
}
which explicitly breaks the $\mathrm{U}(1)$ $p$-form symmetry down to $\mathbb{Z}_N^{}$,
into Eq.~\eqref{brane field action 1}.
Correspondingly, 
the VEV is discretized as 
\aln{\psi[C_p^{}]=\frac{v}{\sqrt{2}}\exp\left(\frac{2\pi ik}{N}\right)~,\quad k=1,2,\cdots,N~,
} 
in a broken phase of $\mathbb{Z}_N$ $p$-form symmetry.
In this case, the phase degrees of freedom,
\aln{\label{eq:psi_A}
\psi[C_p^{}]=\frac{v}{\sqrt{2}}\exp\left(i\int_{C_p^{}}A_{p}^{}\right)~,
}
will no longer be gapless.

The effective theory must be invariant under 
$\mathbb{Z}_N$ $p$-form symmetry corresponding to the shift,
\aln{
A_p^{}~&\rightarrow~A_p^{}+\frac{n}{N}\Lambda_p^{}~,\quad d\Lambda_p^{}=0~,
\label{eq:Ap Zn transform}
}
with $\int_{C_p^{}}\Lambda_p^{}\in2\pi \mathbb{Z}$, where $n\in\mathbb{Z}$. 
For example, when $p=0$, $\int_{C_{p=0}^{}} A_{p=0}^{}=\varphi(x)$ is the periodic scalar field, and it has the periodic potential $V(\varphi)=V(\varphi+2\pi /N)$.
The effective theory must also be invariant under gauge transformation of $A_p$,
\aln{
A_p^{}~\rightarrow~A_p^{}+d\Lambda_{p-1}^{}~,
\label{eq:Ap gauge transform}
}
with $\int_{C_p^{}}d\Lambda_{p-1}\in2\pi \mathbb{Z}$, which is a redundancy in the degrees of freedom of Eq.~\eqref{eq:psi_A}.

\begin{figure}
    \centering
    \includegraphics[scale=0.4]{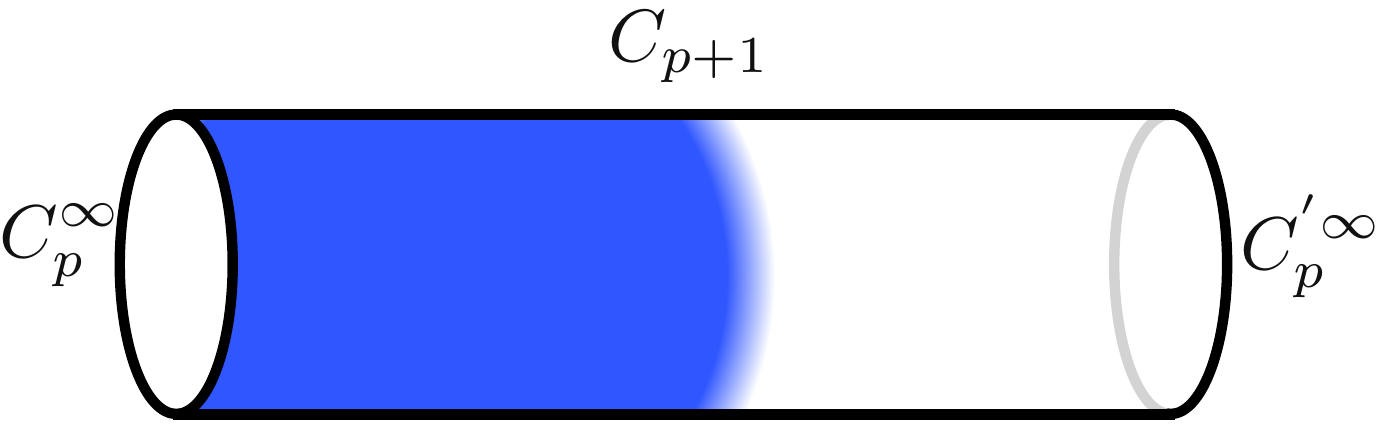}
    \caption{
    $C_{p+1}$ for $p=1$.
    Blue and white colors correspond to different vacua.}
    \label{fig:topological defect}
\end{figure}

When the discrete higher-form symmetry is spontaneously broken, it exhibits topological order. 
The degeneracy of the ground state depends on the topology of the space. 
We assume that the space manifold has a nontrivial topology such as $\Sigma_{d-1}^{}=C_{p+1}^{}\times D_{d-p-2}^{}$, where $C_{p+1}^{}$ is a $(p+1)$-dimensional subspace with boundaries at infinity $C_p^{'\infty}$ and $C_p^{\infty}$,
and $D_{d-p-2}^{}$ is a $(d-p-2)$-dimensional subspace. 
See Fig.~\ref{fig:topological defect} as an example.
%
We also assume that $C_p^{'\infty}$ and $C_p^{\infty}$ are not contractible, so that the $p$-form symmetry can act nontrivially. 
In such a case, there exists a classical static configuration 
$\psi_W^{}[C_p^{}]=v\exp\left(i\int_{C_p^{}}A_{p}^{W}\right)
$
connecting different ground states. 
Note that in the case of $0$-form symmetry, which exhibits not a topological order but a spontaneous breaking of discrete symmetry, the topological defect connecting the different ground states is nothing but a domain wall. 
The corresponding topological charge is given by
\aln{
Q_{d-p-2}^{}
=\frac{N}{2\pi}\int_{C_{p+1}^{}}dA_p^{}=\frac{N}{2\pi}\left(\int_{C_p^{'\infty}}-\int_{C_p^{\infty}}\right)A_p^{}~,
}
More explicitly, $dA_p^W$ can be represented as
\aln{\label{domain wall form}
dA_p^W=\frac{2\pi}{N}\delta_{p+1}(D_{d-p-1}^{W})~,
}
in the thin wall limit, where $\delta_{p+1}(D_{d-p-1}^{W})$ is the Poincar\'e-dual form defined in Eq.~\eqref{eq:Poincare dual form}.
Here, $D_{d-p-1}^W$ corresponds to the worldvolume of $D_{d-p-2}^{}$. 

Now, let us study the low-energy effective theory. 
We generalize the argument of $0$-form symmetry discussed in Ref.~\cite{Hidaka:2019mfm} to the brane field theory.
To derive the effective theory, we rewrite Eq.~\eqref{breaking term} for large $v$ as
\begin{align}\label{eq:Villain action}
    -\lambda_N^{}\left\{(\psi[C_p^{}])^N+(\psi^\dagger [C_p^{}])^N\right\}
    &=-2^{-\frac{N}{2}}\lambda_N^{}{v^N}\cos\left(N\int_{C_p}A_p \right)\notag\\
    &\to 2^{-\frac{N}{2}-1}\lambda_Nv^N
    \left(N\int_{C_p}A_p-2\pi n\right)^2~.
\end{align}
In the last line, we approximated the cosine by using the Villain formula \cite{Villain:1977}, 
\begin{equation}
\exp({\beta\cos\theta})  \approx \sum_{n\in \mathbb{Z}} \exp\left({\beta-\frac{1}{2}\beta(\theta -2\pi n)^2}\right)~,
\end{equation}
for large $\beta$, and dropped the constant term. 
Equation~\eqref{eq:Villain action} is gauge invariant under Eq.~\eqref{eq:Ap gauge transform} accompanied by the shift of $n$, $n\to n+N\int_{C_p}d\Lambda_{p-1}/(2\pi)$. 
Similarly, it is invariant under $\mathbb{Z}_N$ transformation~\eqref{eq:Ap Zn transform}.
We can replace the integer $n$ in Eq.~\eqref{eq:Villain action} by introducing the flat $\mathrm{U}(1)$ gauge field ${f}_p$, 
as $N\int_{C_p}A_p-2\pi n=\int_{C_p}(NA_p-{f}_p)$, where $\int_{C_p}{f}_p\in2\pi\mathbb{Z}$.

By performing the same calculations as in Sec.~\ref{sec:NG mode}, we have the following effective action of $A_p^{}$:
\aln{\label{effective action on defect}
S[A]&=-\frac{v^2}{2}\int_{\Sigma_d^{}}F_{p+1}^{}\wedge \star F_{p+1}^{}-\frac{\tilde{\lambda}_N^{}}{2}\int_{\Sigma_d^{}}(NA_p^{}-f_p)\wedge \star(NA_p^{}-f_p)
-\frac{1}{2\pi}\int_{\Sigma_d}B_{d-p-1}\wedge df_p~,
}
where $\tilde{\lambda}_N^{}$ is a coupling constant which includes $\lambda_N^{}$. 
See Appendix~\ref{sec:mass} for the derivation of the mass term. 
Here, $f_p$ is the $\mathrm{U}(1)$ $p$-form gauge field, and the flatness condition is imposed by the last term by using the Lagrange multiplier $B_{d-p-1}$.\footnote{
Note also that when one considers  a topological defect $A_p^W$ as a background solution, $f_p^{}$ is replaced by $f_p^{}-NA_p^W$ in the last term in Eq.~(\ref{effective action on defect})
}
 On the other hand, $f_p$ can be eliminated using the equation of motion for $f_p^{}$,
\aln{
f_p^{}=NA_p^{}+\frac{(-1)^{d-p}}{2\pi {\tilde{\lambda}}_N^{}}\star dB_{d-p-1}^{}~,
}
which leads to
\aln{
&-\frac{1}{8\pi^2 \tilde{\lambda}_N}\int_{\Sigma_d} dB_{d-p-1}\wedge\star dB_{d-p-1}
+\frac{1}{4\pi^2\tilde{\lambda}_N}\int_{\Sigma_d} d(B_{d-p-1}\wedge\star dB_{d-p-1})\notag\\
&\quad
-\frac{v^2}{2}\int_{\Sigma_d}F_{p+1}^{}\wedge \star F_{p+1}^{}
-\frac{N}{2\pi}\int_{\Sigma_d^{}} B_{d-p-1}^{}\wedge dA_p^{}
~.
}
For the domain wall configuration~(\ref{domain wall form}), the last term becomes 
\begin{equation}
    \frac{N}{2\pi}\int_{\Sigma_d^{}} B_{d-p-1}^{}\wedge dA_p^{} = 
    \int_{D_{d-p-1}^{W}} B_{d-p-1}^{},
\end{equation}
which implies that the worldvolume $D_{d-p-1}^W$ couples with the gauge field $B_{d-p-1}^{}$.

In the low-energy limit, we can neglect higher derivative terms, and we obtain the topological field theory with the action,
\aln{
\label{eq:BF}
S_{\rm top}^{}=-\frac{N}{2\pi}\int_{\Sigma_d^{}} B_{d-p-1}^{}\wedge dA_p^{}~.
}
This effective theory has the following emergent global $\mathbb{Z}_N^{}$ $(d-p-1)$-form symmetry:
\aln{
B_{d-p-1}^{}~&\rightarrow~B_{d-p-1}^{}+\frac{n}{N}\Lambda_{d-p-1}^{}~,\quad d\Lambda_{d-p-1}^{}=0,
\quad n\in\mathbb{Z}~,
}
with
\aln{
\int_{C_{d-p-1}^{}}\Lambda_{d-p-1}^{}\in2\pi \mathbb{Z}~,
}
in addition to the original $\mathbb{Z}_N$ $p$-form symmetry~\eqref{eq:Ap Zn transform},
where $C_{d-p-1}^{}$ is a $(d-p-1)$-dimensional closed subspace. 
The charged objects for $p$- and $(d-p-1)$-form symmetries
are the Wilson surfaces:
\aln{
W[C_{p}]=\exp\left(i\int_{C_p^{}}A_p^{}\right)~,\quad V[C_{d-p-1}]=\exp\left(i\int_{C_{d-p-1}^{}}B_{d-p-1}^{}\right)~,
}
respectively. 
We can also show that $V[C_{d-p-1}^{}]$ and $W[C_p^{}]$ correspond to the symmetry operators of the above $p$ and $(d-p-1)$-form symmetries. 
They satisfy
\aln{
\langle V[C_{d-p-1}^{}]W[C_p^{}]\dots\rangle =e^{\frac{2\pi i}{N} \mathrm{Link}[C_p^{},C_{d-p-1}^{}]}\langle W_q[C_p^{}]\dots\rangle~,
}
where $\mathrm{Link}[C_p^{},C_{d-p-1}^{}]$ is the linking number defined in Eq.~\eqref{eq:Linking number},
and ``$\dots$'' denotes other operators that neither link nor intersect $C_{d-p-1}$.

As mentioned above, this theory~\eqref{eq:BF} exhibits the grand state degeneracy depending on the topology of the spatial manifold  $\Sigma_{d-1}^{}$.
Let us look at this in detail using the same argument in Sec.~\ref{sec:Conservation law}. 
When $\Sigma_{d-1}=S^{p}\times S^{d-p-1}$, we can choose $C_{p}=S^p$ and $C_{d-p-1}^{}=S^{d-p-1}$. 
Consider 
$V[C_{d-p-1}^{}]W[C_p^{}]V^{-1}[C_{d-p-1}^{}]$ in the operator formalism at time $t$. 
The ordering of the operator product corresponds to the time ordering. 
That is, the pair of symmetry operators $V[C_{d-p-1}]$ and $V^{-1}[C_{d-p-1}]$, corresponds to the operator on
$C_{d-p-1}(t+\eta)\cup \overline{C_{d-p-1}}(t-\eta)$ in the path integral formalism.  
Here $\eta$ is an infinitesimal parameter and $\overline{C_{d-p-1}}$ is the subspace with the opposite orientation of ${C}_{d-p-1}(t)$. 
In this case, $C_p(t)$ and $C_{d-p-1}(t+\eta)\cup \overline{C_{d-p-1}}(t-\eta)$ can be linked in space-time. This means that 
\aln{
V[C_{d-p-1}^{}]W[C_p^{}]V^{-1}[C_{d-p-1}^{}]=e^{\frac{2\pi i}{N}}W[C_p^{}]~,
}
holds as an operator relation. 
Since both operators are symmetry operators, we can choose a groundstate $|\Omega \rangle$ as an eigenstate of one of the symmetry operators.
Here, we take the eigenstate of $V[C_{d-p-1}^{}]$, i.e., $V[C_{d-p-1}^{}]|\Omega \rangle=e^{i\theta}|\Omega \rangle$,
where $e^{i\theta}$ is the eigevalue. 
Since $W[C_p^{}]$ is also a symmetry operator,
\aln{
|\Omega'\rangle=W[C_p^{}]|\Omega\rangle~
}
has the same energy as $|\Omega\rangle$. 
But it has a different eigenvalue of $V[C_{d-p-1}^{}]$,
\aln{V[C_{d-p-1}^{}]|\Omega'\rangle=e^{i\theta+\frac{2\pi i}{N}}|\Omega'\rangle~.
}
Since $|\Omega\rangle$ and $|\Omega'\rangle$ have different eigenvalues, they are orthogonal, $\langle \Omega'|\Omega\rangle=0$; 
that is, the ground state is degenerate.

\subsection{Brane field model for superconductor}\label{sec:superconductor}
We here discuss a superconducting phase (Higgs phase) in a brane-field model by coupling $(p+1)$-form gauge field.
We mostly focus on the low-energy degrees of freedom, and leave more detailed studies including the massive degrees of freedom for future investigations.  

We consider a gauged $p$-form brane-field model:
\aln{
S={\cal N}\int [dC_{p}^{}]\left\{-
\int_{\Sigma_d^{}}\frac{\delta(C_{p}^{})}{\mathrm{Vol}[C_{p}^{}]}D_{G}^{}\psi^\dagger\wedge \star D_{G}^{}\psi^{}-V(\psi^\dagger \psi)
\right\}-\frac{1}{2g^2}\int H_{p+2}^{}\wedge \star H_{p+2}^{}~,
\label{toy model}
}
where $g^2$ is a gauge coupling whose mass dimension is $2(p+2)-d$ and 
\aln{D_G^{}\psi[C_p^{}]=D\psi[C_p^{}]-iqB_{p+1}^{}\psi[C_p^{}]~,\quad H_{p+2}^{}=dB_{p+1}^{}~.
}
Here, $B_{p+1}$ is the $\mathrm{U}(1)$ $(p+1)$-form gauge field.
Note that we consider a general charge $q\in\mathbb{Z}$ compared to Eq.~(\ref{covariant derivative})\footnote{
One may think that the charge can always be absorbed into the gauge coupling by the field redefinition $qB_{p+1}^{}\rightarrow B_{p+1}^{}$. 
However, this is not true since such a redefinition changes the Dirac quantization condition $\int_{C_{p+2}^{}}H_{p+2}\in 2\pi \mathbb{Z}$.  
In other words, for a given quantization condition, the charge is determined up to $\mathbb{Z}$. 
}.
The action is invariant under $p$-form gauge transformation,
\begin{equation}
\psi[C_p^{}]\rightarrow e^{iq\int_{C_p^{}}\Lambda_p^{}}
\psi[C_p^{}]~,\quad 
B_{p+1}^{}  \rightarrow  B_{p+1}^{}+d\Lambda_{p}^{}~,
\label{eq:gauge transformation Bp}
\end{equation}
where $\Lambda_p$ is $p$-form normalized as $\int_{C_{p+1}^{}}d\Lambda_{p}^{}\in 2\pi\mathbb{Z}$.
In addition to the $p$-form gauge symmetry, when $q>1$, this theory has a global electric $\mathbb{Z}_q^{}$ $(p+1)$-form symmetry: 
\aln{\label{p+1 form symmetry}
B_{p+1}^{} \rightarrow B_{p+1}^{}+\frac{1}{q}\Lambda_{p+1}^{}~,\quad d\Lambda_{p+1}^{}=0~,\quad  \int_{C_{p+1}^{}}\Lambda_{p+1}^{}\in 2\pi\mathbb{Z}~, 
}
where $C_{p+1}^{}$ is a $(p+1)$-dimensional closed subspace.  
The corresponding symmetry operator and charged objects are
\aln{
U[C_{d-p-2}] &= \exp\left(i\frac{2\pi}{q}\frac{1}{g^2}\int_{C_{d-p-2}}\star H_{p+2}\right)~,\\
W[C_{p+1}^{}]&=\exp\left(i\int_{C_{p+1}^{}}B_{p+1}^{}\right)~,
}
respectively, where $C_{d-p-2}$ is a $(d-p-2)$-dimensional closed subspace. 
The discrete symmetry means that $U[C_{d-p-2}]$ is a topological operator, which can be checked as follows.
By deforming $C_{d-p-2}$ by $C_{d-p-2}+\partial D_{d-p-1}$, we obtain 
\aln{
U[C_{d-p-2}+\partial D_{d-p-1}] &= U[C_{d-p-2}]\exp\left(i\frac{2\pi}{q}\frac{1}{g^2}\int_{\partial D_{d-p-1}}\star H_{p+2}\right)
~,
}
where $D_{d-p-1}$ is a $(d-p-1)$-dimensional subspace with boundary. 
Using the Stokes theorem and the Maxwell equation, $(-1)^pd\star H_{p+2}/g^2=q \star J_{p+1}$, we obtain
\aln{
U[C_{d-p-2}+\partial D_{d-p-1}] &= U[C_{d-p-2}]\exp\left(2\pi i(-1)^p\int_{D_{d-p-1}}\star J_{p+1}\right)
~.
}
Here, $q\star J_{p+1}$ is the gauge current defined by the variation of $B_{p+1}$ in the matter part of the action as $\delta S_\mathrm{matter}=-\int_{\Sigma_d}\delta B_{p+1}\wedge q\star J_{p+1} $. Since the charge is quantized to integer, $\int_{D_{d-p-2}}\star J_{p+1}\in \mathbb{Z}$, we obtain $U[C_{d-p-2}+\partial D_{d-p-1}] =U[C_{d-p-2}]$. 
Therefore, $U[C_{d-p-2}]$ is a topological operator.

In addition, the theory has the magnetic $\mathrm{U}(1)$ $(d-p-3)$-form symmetry, whose charge is $Q_{d-p-3}=\int_{C_{p+2}}H_{p+2}/(2\pi)\in\mathbb{Z}$.
The charged object is $(d-p-3)$-dimensional 't Hooft operator.

In the following, we consider the Higgs phase, i.e., we assume that there exists a nontrivial minimum $\psi[C_p^{}]=v/\sqrt{2}$ in the potential $V(\psi^\dagger\psi)$. 
In order to study the low-energy effective theory in a Higgs phase, we focus on the phase modulation in the brane field:~\footnote{As mentioned before, the system generally contains many other fluctuations.
However, they typically become massive unless they are protected by other symmetries that forbid their mass terms. 
}
\aln{
\psi[C_{p}^{}]=\frac{v}{\sqrt{2}}\exp\left(
i\int_{C_p^{}} A_{p}^{}\right)~.
}
Then, by repeating the same calculations as before, Eq.~(\ref{toy model}) becomes 
\aln{\label{dual abelian Higgs}
\int_{\Sigma_d^{}}&\left[-\frac{1}{2g^2}H_{p+2}^{}\wedge \star H_{p+2}^{}-\frac{\lambda}{2(2\pi)}(F_{p+1}^{}-q_{}^{}B_{p+1}^{})\wedge \star (F_{p+1}^{}-q_{}^{}B_{p+1}^{})\right]~,
}
where $\lambda$ is a parameter whose mass dimension is $d-2(p+1)$, and $F_{p+1}^{}=dA_{p}^{}$.  
In addition to the original $p$-form gauge symmetry~\eqref{eq:gauge transformation Bp},  this effective theory has a $(p-1)$-form gauge symmetry given by
\aln{
A_p^{}\quad &\rightarrow \quad A_p^{}+d\Lambda_{p-1}^{}~,\quad \int_{C_p^{}}d\Lambda_{p-1}^{} \in 2\pi\mathbb{Z}~.
}
Equation~(\ref{dual abelian Higgs}) corresponds to the low-energy effective action of the Abelian-Higgs model in the broken phase~\cite{Banks:2010zn,Hidaka:2019jtv}. 
This effective theory has an emergent $\mathrm{U}(1)$ $(d-p-2)$-form symmetry, whose charge is given by 
\aln{
Q_{d-p-2}=\frac{1}{ 2\pi}\int_{C_{p+1}^{}}F_{p+1}^{}\in \mathbb{Z}~,
}
where $C_{p+1}^{}$ is a $(p+1)$-dimensional closed subspace.  
The charged object is the $(d-p-2)$-dimensional 't Hooft operator, which is a defect operator formally obtained by excising a codimension $(p+2)$ dimensional locus from $\Sigma_d$ and imposing a boundary condition on $A_p$ around it.  
Instead, one can express the 't Hooft operator by using a field in the dualized theory. 
By introducing the dual field of $A_p^{}$ as $\tilde{A}_{d-p-2}^{}$, Eq.~(\ref{dual abelian Higgs}) can be dualized as
\aln{ 
\label{dual abelian Higgs 1}
\int_{\Sigma_d^{}}\left[
-\frac{1}{2g^2}H_{p+2}^{}\wedge \star H_{p+2}^{}-\frac{q_{}^{}}{2\pi}B_{p+1}^{}\wedge \tilde{F}_{d-p-1}^{}-\frac{1}{2(2\pi)\Lambda}\tilde{F}_{d-p-1}^{}\wedge \star \tilde{F}_{d-p-1}^{}\right]~,
}
where $\tilde{F}_{d-p-1}=d\tilde{A}_{d-p-2}$ (See Appendix \ref{sec:dualize} for the derivation).
In the dualized theory, the $(d-p-2)$-form symmetry is given by a transformation of $\tilde{A}_{d-p-2}$  as
\aln{
\tilde{A}_{d-p-2}^{}~\rightarrow~\tilde{A}_{d-p-2}^{}+\frac{1}{q}\tilde{\Lambda}_{d-p-2}^{}~,\quad d \tilde{\Lambda}_{d-p-2}^{}=0~,\quad \int_{{C}_{d-p-2}}\tilde{\Lambda}_{d-p-2}^{}\in2\pi\mathbb{Z}~,
}
where $C_{d-p-2}^{}$ is a $(d-p-2)$-dimensional closed subspace. 
The corresponding charge and charged object for the $(d-p-2)$-form symmetry are 
\aln{
Q_{d-p-2} &= \frac{1}{2\pi\Lambda} \int_{C_{p+1}} \star \tilde{F}_{d-p-1}~,\\
V[C_{d-p-2}^{}]&=\exp\left(i\int_{C_{d-p-2}^{}}\tilde{A}_{d-p-2}^{}\right)~,
}
respectively. 
Note that there is a correspondence between the dual theory and original theory, $\star\tilde{F}_{d-p-1}/\Lambda =dF_{p+1}$.

For example, for $d=4$ and $p=0$, the brane field theory (\ref{toy model}) is nothing but the usual Abelian Higgs model, and $W[C_1^{}]$ is the Wilson loop while $V[C_2^{}]$ is a $2$-dimensional surface operator which corresponds to the world surface of the vortex.
%
On the other hand, we can also derive the same effective theory from the brane-field theory with  $d=4$, $p=1$,
where the roles of $B$ and $\tilde{A}$ are reversed.
%
In this theory, the dual gauge field $\tilde{B}_1^{}$ of the original Abelian-Higgs model appears as a phase d.o.f $\psi[C_1^{}]\sim \exp\left(i\int_{C_1^{}}\tilde{B}_1^{}\right)$ which corresponds to the 't Hooft operator for the Abelian-Higgs model. 
More generally, one can see that the gauged $(d-p-3)$-form brane field theory gives the same-low-energy effective theory as Eq.~(\ref{dual abelian Higgs 1}) and that the roles of scalar and gauge fields are exchanged each other.  
\section{Summary and discussion}\label{sec:summary}
We have proposed an effective brane field theory with higher-form symmetry by generalizing the previous work for a mean string field theory~\cite{Iqbal:2021rkn}.
As a generalization of the ordinary field $\phi(x)$ for $p=0$, the fundamental field $\psi[C_p^{}]$ that is charged under the $p$-form transformation is defined as a functional of the $p$-dimensional brane $C_p^{}$.
We constructed an action that is invariant under the higher-form transformation using the area derivative acting on higher-dimensional objects.
Furthermore, we have discussed the spontaneous breaking of both $\mathrm{U}(1)$ and discrete higher-form symmetries and studied their low-energy effective theories, which are $p$-form Maxwell and the BF-type topological field theories, respectively. 

There are several issues to be addressed.  
First, while we have focused on closed subspaces in this paper, we can generalize to branes with boundaries. 
In this case, the area derivatives need to be treated carefully since we have contributions from both the bulk and the boundary.
Compared to the closed-manifold case, 
one of the crucial differences is that low-energy effective theory typically contains other higher-form fields originating from the boundary d.o.f as well as the bulk ones. 
Such an effective theory might have emergent gauge symmetry as well as emergent higher-form global symmetry. 
%
%
%

Second, we have considered an effective theory for a single type of extended object, but it would be interesting to consider a theory in which objects of different dimensions interact.
Additionally, a theory that includes objects constrained on an extended object or on the intersection of extended objects can also be considered.
Symmetries of such a theory could be described by higher groups or, more generally, non-invertible symmetries.
It is possible that a theory exhibits anomalies where symmetries are broken by quantum corrections. It would be interesting to consider whether an anomaly specific to brane field theory could exist.

Finally, it is interesting to study a brane field theory without Lorentz invariance.  
In the case of $0$-form symmetry without Lorentz invariance, there exist two types of Nambu-Goldstone modes, and unlike in Lorentz-invariant systems, there is no one-to-one correspondence between the generators of the broken symmetry and the Nambu-Goldstone modes~\cite{Nielsen:1975hm,Schafer:2001bq,Miransky:2001tw,Nambu:2004yia,Watanabe:2011ec}. 
The complete relation can be understood by considering the expectation value of the commutation relation of broken generators~\cite{Watanabe:2012hr,Hidaka:2012ym,Watanabe:2014fva,Hayata:2014yga}. 
This concept has been extended to the case with $p$-form symmetry without Lorentz invariance using a low-energy effective theory~\cite{Hidaka:2020ucc}. 
It is interesting to study how the low-energy effective theory is derived from the perspective of the brane field theory.

%

%
%
%

We would like to investigate these problems in our future work.

\section*{Acknowledgements}
Y.H. would like to thank Ryo Yokokura for the useful discussions.
The work of K.K. is supported by KIAS Individual Grants, Grant No. 090901.  
The work of Y.H. is supported by Japan Society for the Promotion of Science (JSPS) KAKENHI Grant Nos. 21H01084.

\appendix

\section{Differential forms}\label{differential forms}
We summarize the basics of differential forms. 
We consider a $d$-dimensional spacetime $\Sigma_d^{}$.
The totally anti-symmetric tensor is represented by $\epsilon_{\mu_1^{}\cdots \mu_d^{}}^{}$. 
In particular, we have
\aln{
\epsilon^{\mu_1^{}\cdots \mu_d^{}}=g^{\mu_1^{}\nu_1^{}}\cdots g^{\mu_d^{}\nu_d^{}}\epsilon_{\nu_1^{}\cdots \nu_r^{}}^{}=g^{-1}\epsilon_{\mu_1^{}\cdots \mu_d^{}}^{}~.
}
We also define 
\aln{\eta_{\mu_1^{}\cdots \mu_d^{}}^{}\coloneqq\sqrt{-g}\epsilon_{\mu_1^{}\cdots \mu_d^{}}^{}\quad \leftrightarrow\quad \eta^{\mu_1^{}\cdots \mu_d^{}}=\frac{1}{\sqrt{-g}}\epsilon_{\mu_1^{}\cdots \mu_d^{}}^{}~.
}
On a $p$-dimensional subspace $C_p^{}$, we have
\aln{\eta_{i_1^{}\cdots i_p^{}}^{}\coloneqq\sqrt{h}\epsilon_{i_1^{}\cdots i_p^{}}^{}\quad \leftrightarrow\quad \eta^{i_1^{}\cdots i_p^{}}=\frac{1}{\sqrt{h}}\epsilon_{i_1^{}\cdots i_p^{}}^{}~.
\label{totally anti-symmetric tensor}
}
Let 
\aln{\omega_p^{}=\frac{1}{p!}\omega_{\mu_1^{}\cdots \mu_{p}^{}}dX^{\mu_1^{}}\wedge \cdots \wedge dX^{\mu_p^{}}~
}
be a general $p$-form.  
Then, the Hodge dual is defined by 
\aln{\star \omega_p^{}=\frac{\sqrt{-g}}{p!(d-p)!}\omega_{\mu_1^{}\cdots \mu_{p}^{}}{\epsilon^{\mu_1^{}\cdots \mu_p^{}}}_{\nu_1^{}\cdots \nu_{d-p}^{}}dX^{\nu_1^{}}\wedge \cdots \wedge dX^{\nu_{d-p}^{}}~.
}
For a Lorentzian spacetime $\Sigma_d^{}$, we have
\aln{\star \star \omega_p^{}=(-1)^{1+p(d-p)}\omega_p^{}~.
}
As usual, we can construct the integral over $\Sigma_d^{}$ by
\aln{\int_{\Sigma_d^{}}\omega_p^{} \wedge \star \omega_p^{}~.
}
However, what we want is an integration over $C_p^{}$. 
To construct it, we define 
\aln{\delta(C_p^{})\coloneqq\int_{C_p^{}}^{} d^p\xi\sqrt{-\frac{h}{g}}\prod_{\mu=0}^{d-1}\delta\left(X^\mu-X^\mu(\xi)\right)~,
}
which leads to
\aln{\int_{\Sigma_d^{}}^{}\delta(C_p^{})\omega_p^{} \wedge \star \omega_p^{}=\frac{1}{p!}\int_{C_p^{}} d^p\xi\sqrt{h}\omega_{\mu_1^{}\cdots \mu_p^{}}^{}(X(\xi))\omega^{\mu_1^{}\cdots \mu_p^{}}(X(\xi))~.
}
%

\section{Truncated action}\label{sec:truncation}

When the brane field $\psi[C_p^{}]$ is given by a functional as Eq.~(\ref{functional form}), the action in Eq.~(\ref{brane field action 1}) becomes 
\aln{
S_0^{}[\{A_p^{(a)}\}]=-{\cal N}\int [dC_p^{}]\left(\frac{1}{\mathrm{Vol}[C_p^{}]}\sum_{a,b}\int_{\Sigma_d^{}}\delta(C_p^{})
F^{(a)}_{p+1}\wedge \star F^{(b)}_{p+1}
\frac{\partial \psi^\dagger}{\partial z^a}\frac{\partial \psi}{\partial z^b}+V(\psi^\dagger \psi)
\right)~.
}
By inserting 
\aln{
1=\prod_a \left(\int dz_a^{}\delta\left(\int_{C_p^{}}A_p^{(a)}-z_a^{}\right)\right)~,
} we have
\aln{
S_0^{}[\{A_p^{(a)}\}]=\int\left(\prod_a dz_a^{} \right)\left(-g^{ab}(z)\frac{\partial \psi^\dagger}{\partial z^a}\frac{\partial \psi}{\partial z^b}-g(z)V(\psi^\dagger \psi)\right)~,
\label{emergent QFT}
}
where
\aln{
g^{ab}(z)&={\cal N}\int [dC_p^{}]\frac{1}{\mathrm{Vol}[C_p^{}]}\int_{\Sigma_d^{}}\delta(C_p^{})
F^{(a)}_{p+1}\wedge \star F^{(b)}_{p+1}
\prod_{a} \delta\left(\int_{C_p^{}}A_p^{(a)}-z_a^{}\right)~,
\label{metric 1}
\\
g(z)&={\cal N}\int [dC_p^{}]\prod_a\delta\left(\int_{C_p^{}}A_p^{(a)}-z_a^{}\right)~.
\label{metric 2}
}
The truncated action in Eq.~(\ref{emergent QFT}) can be interpreted as 
a field theory on a curved manifold, whose background metric is determined by the brane configurations in Eqs.~(\ref{metric 1}) and (\ref{metric 2}). 

\section{Calculation of mass term}\label{sec:mass}
The effective action for broken $\mathbb{Z}_q$ $p$-form symmetry discussed in Sec.~\eqref{sec:Discrete} contains  
\aln{
&{\cal N}\int [dC_p^{}]\left(\int_{C_p^{}} (NA_p^{}-f_p^{})\right)^2\notag\\
&={\cal N}\int [dC_p^{}]\left(\frac{1}{p!}\int d^p\xi\sqrt{h} E^{\mu_1^{}\cdots \mu_p^{}}(X(\xi))\delta A_{p,\mu_1^{}\cdots \mu_p^{}}^{}(X(\xi))\right)^2~,
}
where $\delta A_p^{}=NA_p^{}-f_p^{}$. 
Following the same procedures as Sec.~\ref{sec:NG mode}, this can be estimated as
\aln{
&{\cal N}\int [dC_p^{}]\left(\int_{C_p^{}} (NA_p^{}-f_p)\right)^2\notag\\
&\quad=\int \frac{d^dk}{(2\pi)^d}\delta \tilde{A}_{p,\mu_1^{}\cdots \mu_p^{}}^{}(k)\int \frac{d^dk'}{(2\pi)^d}\delta \tilde{A}_{p,\nu_1^{}\cdots \nu_p^{}}^{}(k')\int d^dxe^{i(k_\mu^{}+k'_\mu)x^\mu}\langle E^{\mu_1^{}\cdots \mu_p^{}}E^{\nu_1^{}\cdots \nu_p^{}}\rangle\notag
\\
&\quad=\int \frac{d^dk}{(2\pi)^d}\delta \tilde{A}_{p,\mu_1^{}\cdots \mu_p^{}}^{*}(k)\delta \tilde{A}_{p,\nu_1^{}\cdots \nu_p^{}}^{}(k)\langle E^{\mu_1^{}\cdots \mu_p^{}}E^{\nu_1^{}\cdots \nu_p^{}}\rangle~,
}
where
\aln{
\langle E^{\mu_1^{}\cdots \mu_p^{}}E^{\nu_1^{}\cdots \nu_p^{}}\rangle &={\cal N}\int {\cal D}X_\mathrm{NZ}^{}\frac{1}{\mathrm{Vol}[C_p^{}]}\int_{S_p^{}}d^p\xi \sqrt{h(\xi)}\int_{S_p^{}}d^p\xi' \sqrt{h(\xi')}\notag\\
&\qquad \times e^{-T_p^{}\mathrm{Vol}[C_p^{}]+ik_\mu^{}(X_\mathrm{NZ}^\mu(\xi)-X_\mathrm{NZ}^\mu(\xi'))}
E^{\mu_1^{}\cdots \mu_p^{}}(X_\mathrm{NZ}^{}(\xi))E^{\nu_1^{}\cdots \nu_p^{}}(X_\mathrm{NZ}^{}(\xi'))~.
}
Assuming the spacetime symmetry is not broken, we have 
\aln{
\langle E^{\mu_1^{}\cdots \mu_p^{}}E^{\nu_1^{}\cdots \nu_p^{}}\rangle &=\frac{1}{{(p!)^3}}\sum_{\sigma,\sigma'\in S_p^{}}\mathrm{sgn}(\sigma)\mathrm{sgn}(\sigma') \bigg[c_0^{}(k^2)\eta^{\mu_{\sigma(1)}^{}\nu_{\sigma'(1)}^{}}\cdots \eta^{\mu_{\sigma(p)}^{}\nu_{\sigma'(p)}^{}}
\nn
&\qquad+c_1^{} (k^2)k^{\mu_{\sigma(1)}^{}}k^{\nu_{\sigma'(1)}^{}}\eta^{\mu_{\sigma(2)}^{}\nu_{\sigma'(2)}^{}}\cdots \eta^{\mu_{\sigma(p)}^{}\nu_{\sigma'(p)}^{}}+\cdots\bigg]~,
\label{k expansion}
}
where $c_0^{}(k^2)$ and $c_1^{}(k^2)$ are functions of $k^2$ in general. 
In the low-energy limit, however, we can neglect the $k$ dependence, and the first term gives 
\aln{\int \frac{d^dk}{(2\pi)^d}\left[\frac{c_0^{}(0)}{p!}({\delta\tilde{A}_{p}}^{\mu_1^{}\cdots \mu_p^{}}(k))^*\delta \tilde{A}_{p,\nu_1^{}\cdots \nu_p^{}}^{}(k')
\right]=c_0^{}(0)\int_{\Sigma_d^{}}(NA_p^{}-f_p)\wedge \star (NA_p^{}-f_p)~,
}
which corresponds to the mass term in Eq.~(\ref{effective action on defect}). 

\section{\texorpdfstring{Equivalence between Eqs.~\eqref{dual abelian Higgs} and \eqref{dual abelian Higgs 1}}{Equivalence between Eqs. (135) and (138)}}\label{sec:dualize}

Here, we show the equivalence between Eqs.~\eqref{dual abelian Higgs} and \eqref{dual abelian Higgs 1}. 
We begin with the following ``parent'' action that generates both Eqs.~\eqref{dual abelian Higgs} and \eqref{dual abelian Higgs 1},
\aln{
&\int_{\Sigma_d^{}}\Bigl[
-\frac{1}{2(2\pi)\lambda}(\tilde{F}_{d-p-1}^{}-G_{d-p-1}^{})\wedge \star (\tilde{F}_{d-p-1}^{}-G_{d-p-1}^{})\notag\\
&\qquad-\frac{q_{}^{}}{2\pi}B_{p+1}^{}\wedge (\tilde{F}_{d-p-1}^{}-G_{d-p-1}^{})+\frac{(-1)^{p}}{2\pi}A_{p}^{}\wedge dG_{d-p-1}^{}
\Bigr]~,
\label{BF lagrangian 1}
}
where $G_{d-p-1}$, $B_{p+1}$, $A_p$ are $(d-p-1)$, $(p+1)$, and $p$-form gauge fields, respectively.
$\tilde{F}_{d-p-1}=d\tilde{A}_{d-p-2}^{}$ is the field strength of the $(d-p-2)$-form gauge field $\tilde{A}_{d-p-2}^{}$.

From the the equation of motion of $A_{p}^{}$, we have $dG_{d-p-1}^{}=0$, so $G_{d-p-1}^{}$ can be locally expressed as the exact form $G_{d-p-1}=d\tilde{G}_{d-p-2}^{}$. 
This implies that $\tilde{G}_{d-p-2}^{}$ can be absorbed into the definition of $A_{d-p-2}^{}$ and Eq.~\eqref{BF lagrangian 1} reduces to Eq.~\eqref{dual abelian Higgs 1} except the kinetic term of $B_{p+1}$,
\aln{
\int_{\Sigma_d^{}}\Bigl[
-\frac{1}{2(2\pi)\lambda}\tilde{F}_{d-p-1}^{}\wedge \star \tilde{F}_{d-p-1}^{}
-\frac{q_{}^{}}{2\pi}B_{p+1}^{}\wedge \tilde{F}_{d-p-1}^{}
\Bigr]~.
\label{BF lagrangian 2}
}
On the other hand, if we redefine $G_{d-p-1}^{}$ as $\tilde{F}_{d-p-1}^{}-G_{d-p-1}^{}\rightarrow G_{d-p-1}^{}~$,
the action becomes 
\aln{
\label{eq:reduced action}
\int_{\Sigma_d^{}}\Bigl[-\frac{1}{2(2\pi)\lambda}G_{d-p-1}^{}\wedge \star G_{d-p-1}^{}
+\frac{1}{2\pi}\left(dA_{p}^{}-q_{}^{}B_{p+1}^{}\right)\wedge G_{d-p-1}^{}
\Bigl]
~.
}
Here, we have performed integration by parts for $G_{d-p-1}$.
The equation of motion for $G_{d-p-1}$ is 
\begin{equation}
    G_{d-p-1} =-\lambda \star(dA_{p}-qB_{p+1})~.
    \label{eq:solution of G}
\end{equation}
Inserting Eq.~\eqref{eq:solution of G} into Eq.~\eqref{eq:reduced action}, the action reduces to 
\begin{equation}
\int_{\Sigma_d^{}}\frac{-\lambda}{2(2\pi)}
 (dA_{p}-qB_{p+1})\wedge \star(dA_{p}-qB_{p+1})
~,
\end{equation}
which coincides with Eq.~(\ref{dual abelian Higgs}) except the kinetic term of $B_{p}$. 
One can see that $2\pi \lambda\coloneqq e^2$ corresponds to the gauge coupling and reproduces the same normalization as in Ref.~\cite{Banks:2010zn} for $d=4$.

%
%
%
%
%


%

\bibliographystyle{TitleAndArxiv}
\bibliography{Bibliography}

\providecommand{\bysame}{\leavevmode\hbox to3em{\hrulefill}\thinspace}
\begin{thebibliography}{10}

\bibitem{Landau:1937obd}
L.~D. Landau, \emph{{On the theory of phase transitions}}, Zh. Eksp. Teor. Fiz.
  \textbf{7} (1937), 19--32.

\bibitem{landau2013statistical}
L.~Landau and E.~Lifshitz, \emph{Statistical physics: Volume 5}, no.~5,
  Elsevier Science, 2013.

\bibitem{Gaiotto:2014kfa}
D.~Gaiotto, A.~Kapustin, N.~Seiberg, and B.~Willett, \emph{{Generalized Global
  Symmetries}}, JHEP \textbf{02} (2015), 172,  \texttt{1412.5148}.

\bibitem{Kapustin:2005py}
A.~Kapustin, \emph{{Wilson-'t Hooft operators in four-dimensional gauge
  theories and S-duality}}, Phys. Rev. D \textbf{74} (2006), 025005,
  \texttt{hep-th/0501015}.

\bibitem{Pantev:2005zs}
T.~Pantev and E.~Sharpe, \emph{{GLSM's for Gerbes (and other toric stacks)}},
  Adv. Theor. Math. Phys. \textbf{10} (2006), no.~1, 77--121,
  \texttt{hep-th/0502053}.

\bibitem{Nussinov:2009zz}
Z.~Nussinov and G.~Ortiz, \emph{{A symmetry principle for topological quantum
  order}}, Annals Phys. \textbf{324} (2009), 977--1057,
  \texttt{cond-mat/0702377}.

\bibitem{Banks:2010zn}
T.~Banks and N.~Seiberg, \emph{{Symmetries and Strings in Field Theory and
  Gravity}}, Phys. Rev. D \textbf{83} (2011), 084019,  \texttt{1011.5120}.

\bibitem{Kapustin:2013uxa}
A.~Kapustin and R.~Thorngren, \emph{{Higher symmetry and gapped phases of gauge
  theories}},  (2013),  \texttt{1309.4721}.

\bibitem{Aharony:2013hda}
O.~Aharony, N.~Seiberg, and Y.~Tachikawa, \emph{{Reading between the lines of
  four-dimensional gauge theories}}, JHEP \textbf{08} (2013), 115,
  \texttt{1305.0318}.

\bibitem{Kapustin:2014gua}
A.~Kapustin and N.~Seiberg, \emph{{Coupling a QFT to a TQFT and Duality}}, JHEP
  \textbf{04} (2014), 001,  \texttt{1401.0740}.

\bibitem{Gaiotto:2017yup}
D.~Gaiotto, A.~Kapustin, Z.~Komargodski, and N.~Seiberg, \emph{{Theta, Time
  Reversal, and Temperature}}, JHEP \textbf{05} (2017), 091,
  \texttt{1703.00501}.

\bibitem{Hirono:2018fjr}
Y.~Hirono and Y.~Tanizaki, \emph{{Quark-Hadron Continuity beyond the
  Ginzburg-Landau Paradigm}}, Phys. Rev. Lett. \textbf{122} (2019), no.~21,
  212001,  \texttt{1811.10608}.

\bibitem{Hidaka:2019jtv}
Y.~Hidaka, Y.~Hirono, M.~Nitta, Y.~Tanizaki, and R.~Yokokura,
  \emph{{Topological order in the color-flavor locked phase of a ( 3+1
  )-dimensional U(N) gauge-Higgs system}}, Phys. Rev. D \textbf{100} (2019),
  no.~12, 125016,  \texttt{1903.06389}.

\bibitem{Hidaka:2022blq}
Y.~Hidaka and D.~Kondo, \emph{{Emergent higher-form symmetry in Higgs phases
  with superfluidity}},  (2022),  \texttt{2210.11492}.

\bibitem{Sharpe:2015mja}
E.~Sharpe, \emph{{Notes on generalized global symmetries in QFT}}, Fortsch.
  Phys. \textbf{63} (2015), 659--682,  \texttt{1508.04770}.

\bibitem{Tachikawa:2017gyf}
Y.~Tachikawa, \emph{{On gauging finite subgroups}}, SciPost Phys. \textbf{8}
  (2020), no.~1, 015,  \texttt{1712.09542}.

\bibitem{Cordova:2018cvg}
C.~C\'ordova, T.~T. Dumitrescu, and K.~Intriligator, \emph{{Exploring 2-Group
  Global Symmetries}}, JHEP \textbf{02} (2019), 184,  \texttt{1802.04790}.

\bibitem{Benini:2018reh}
F.~Benini, C.~C\'ordova, and P.-S. Hsin, \emph{{On 2-Group Global Symmetries
  and their Anomalies}}, JHEP \textbf{03} (2019), 118,  \texttt{1803.09336}.

\bibitem{Tanizaki:2019rbk}
Y.~Tanizaki and M.~\"Unsal, \emph{{Modified instanton sum in QCD and
  higher-groups}}, JHEP \textbf{03} (2020), 123,  \texttt{1912.01033}.

\bibitem{DelZotto:2020sop}
M.~Del~Zotto and K.~Ohmori, \emph{{2-Group Symmetries of 6D Little String
  Theories and T-Duality}}, Annales Henri Poincare \textbf{22} (2021), no.~7,
  2451--2474,  \texttt{2009.03489}.

\bibitem{Hidaka:2020iaz}
Y.~Hidaka, M.~Nitta, and R.~Yokokura, \emph{{Higher-form symmetries and 3-group
  in axion electrodynamics}}, Phys. Lett. B \textbf{808} (2020), 135672,
  \texttt{2006.12532}.

\bibitem{Hidaka:2020izy}
Y.~Hidaka, M.~Nitta, and R.~Yokokura, \emph{{Global 3-group symmetry and 't
  Hooft anomalies in axion electrodynamics}}, JHEP \textbf{01} (2021), 173,
  \texttt{2009.14368}.

\bibitem{Brennan:2020ehu}
T.~D. Brennan and C.~Cordova, \emph{{Axions, higher-groups, and emergent
  symmetry}}, JHEP \textbf{02} (2022), 145,  \texttt{2011.09600}.

\bibitem{Hidaka:2021mml}
Y.~Hidaka, M.~Nitta, and R.~Yokokura, \emph{{Topological axion electrodynamics
  and 4-group symmetry}}, Phys. Lett. B \textbf{823} (2021), 136762,
  \texttt{2107.08753}.

\bibitem{Hidaka:2021kkf}
Y.~Hidaka, M.~Nitta, and R.~Yokokura, \emph{{Global 4-group symmetry and
  \textquoteright{}t Hooft anomalies in topological axion electrodynamics}},
  PTEP \textbf{2022} (2022), no.~4, 04A109,  \texttt{2108.12564}.

\bibitem{Apruzzi:2021mlh}
F.~Apruzzi, L.~Bhardwaj, D.~S.~W. Gould, and S.~Schafer-Nameki, \emph{{2-Group
  symmetries and their classification in 6d}}, SciPost Phys. \textbf{12}
  (2022), no.~3, 098,  \texttt{2110.14647}.

\bibitem{Barkeshli:2022edm}
M.~Barkeshli, Y.-A. Chen, P.-S. Hsin, and R.~Kobayashi, \emph{{Higher-group
  symmetry in finite gauge theory and stabilizer codes}},  (2022),
  \texttt{2211.11764}.

\bibitem{Nakajima:2022feg}
T.~Nakajima, T.~Sakai, and R.~Yokokura, \emph{{Higher-group structure in
  2n-dimensional axion-electrodynamics}}, JHEP \textbf{01} (2023), 150,
  \texttt{2211.13861}.

\bibitem{Radenkovic:2022qkd}
T.~Radenkovic and M.~Vojinovic, \emph{{Topological invariant of 4-manifolds
  based on a 3-group}}, JHEP \textbf{07} (2022), 105,  \texttt{2201.02572}.

\bibitem{Bhardwaj:2022scy}
L.~Bhardwaj and D.~S.~W. Gould, \emph{{Disconnected 0-form and 2-group
  symmetries}}, JHEP \textbf{07} (2023), 098,  \texttt{2206.01287}.

\bibitem{Kan:2023yhz}
N.~Kan, O.~Morikawa, Y.~Nagoya, and H.~Wada, \emph{{Higher-group structure in
  lattice Abelian gauge theory under instanton-sum modification}}, Eur. Phys.
  J. C \textbf{83} (2023), no.~6, 481,  \texttt{2302.13466}.

\bibitem{Bhardwaj:2017xup}
L.~Bhardwaj and Y.~Tachikawa, \emph{{On finite symmetries and their gauging in
  two dimensions}}, JHEP \textbf{03} (2018), 189,  \texttt{1704.02330}.

\bibitem{Chang:2018iay}
C.-M. Chang, Y.-H. Lin, S.-H. Shao, Y.~Wang, and X.~Yin, \emph{{Topological
  Defect Lines and Renormalization Group Flows in Two Dimensions}}, JHEP
  \textbf{01} (2019), 026,  \texttt{1802.04445}.

\bibitem{Ji:2019jhk}
W.~Ji and X.-G. Wen, \emph{{Categorical symmetry and noninvertible anomaly in
  symmetry-breaking and topological phase transitions}}, Phys. Rev. Res.
  \textbf{2} (2020), no.~3, 033417,  \texttt{1912.13492}.

\bibitem{Komargodski:2020mxz}
Z.~Komargodski, K.~Ohmori, K.~Roumpedakis, and S.~Seifnashri, \emph{{Symmetries
  and strings of adjoint QCD$_{2}$}}, JHEP \textbf{03} (2021), 103,
  \texttt{2008.07567}.

\bibitem{Nguyen:2021yld}
M.~Nguyen, Y.~Tanizaki, and M.~\"Unsal, \emph{{Semi-Abelian gauge theories,
  non-invertible symmetries, and string tensions beyond $N$-ality}}, JHEP
  \textbf{03} (2021), 238,  \texttt{2101.02227}.

\bibitem{Heidenreich:2021xpr}
B.~Heidenreich, J.~McNamara, M.~Montero, M.~Reece, T.~Rudelius, and
  I.~Valenzuela, \emph{{Non-invertible global symmetries and completeness of
  the spectrum}}, JHEP \textbf{09} (2021), 203,  \texttt{2104.07036}.

\bibitem{Koide:2021zxj}
M.~Koide, Y.~Nagoya, and S.~Yamaguchi, \emph{{Non-invertible topological
  defects in 4-dimensional $\mathbb {Z}_2$ pure lattice gauge theory}}, PTEP
  \textbf{2022} (2022), no.~1, 013B03,  \texttt{2109.05992}.

\bibitem{Kaidi:2021xfk}
J.~Kaidi, K.~Ohmori, and Y.~Zheng, \emph{{Kramers-Wannier-like Duality Defects
  in (3+1)D Gauge Theories}}, Phys. Rev. Lett. \textbf{128} (2022), no.~11,
  111601,  \texttt{2111.01141}.

\bibitem{Choi:2021kmx}
Y.~Choi, C.~Cordova, P.-S. Hsin, H.~T. Lam, and S.-H. Shao,
  \emph{{Noninvertible duality defects in 3+1 dimensions}}, Phys. Rev. D
  \textbf{105} (2022), no.~12, 125016,  \texttt{2111.01139}.

\bibitem{Roumpedakis:2022aik}
K.~Roumpedakis, S.~Seifnashri, and S.-H. Shao, \emph{{Higher Gauging and
  Non-invertible Condensation Defects}}, Commun. Math. Phys. \textbf{401}
  (2023), no.~3, 3043--3107,  \texttt{2204.02407}.

\bibitem{Bhardwaj:2022yxj}
L.~Bhardwaj, L.~E. Bottini, S.~Schafer-Nameki, and A.~Tiwari,
  \emph{{Non-invertible higher-categorical symmetries}}, SciPost Phys.
  \textbf{14} (2023), no.~1, 007,  \texttt{2204.06564}.

\bibitem{Cordova:2022ieu}
C.~Cordova and K.~Ohmori, \emph{{Noninvertible Chiral Symmetry and Exponential
  Hierarchies}}, Phys. Rev. X \textbf{13} (2023), no.~1, 011034,
  \texttt{2205.06243}.

\bibitem{Bashmakov:2022jtl}
V.~Bashmakov, M.~Del~Zotto, and A.~Hasan, \emph{{On the 6d origin of
  non-invertible symmetries in 4d}}, JHEP \textbf{09} (2023), 161,
  \texttt{2206.07073}.

\bibitem{Choi:2022rfe}
Y.~Choi, H.~T. Lam, and S.-H. Shao, \emph{{Noninvertible Time-Reversal
  Symmetry}}, Phys. Rev. Lett. \textbf{130} (2023), no.~13, 131602,
  \texttt{2208.04331}.

\bibitem{Bartsch:2022mpm}
T.~Bartsch, M.~Bullimore, A.~E.~V. Ferrari, and J.~Pearson,
  \emph{{Non-invertible Symmetries and Higher Representation Theory I}},
  (2022),  \texttt{2208.05993}.

\bibitem{Apruzzi:2022rei}
F.~Apruzzi, I.~Bah, F.~Bonetti, and S.~Schafer-Nameki, \emph{{Noninvertible
  Symmetries from Holography and Branes}}, Phys. Rev. Lett. \textbf{130}
  (2023), no.~12, 121601,  \texttt{2208.07373}.

\bibitem{GarciaEtxebarria:2022vzq}
I.~n. Garc\'\i{}a~Etxebarria, \emph{{Branes and Non-Invertible Symmetries}},
  Fortsch. Phys. \textbf{70} (2022), no.~11, 2200154,  \texttt{2208.07508}.

\bibitem{Niro:2022ctq}
P.~Niro, K.~Roumpedakis, and O.~Sela, \emph{{Exploring non-invertible
  symmetries in free theories}}, JHEP \textbf{03} (2023), 005,
  \texttt{2209.11166}.

\bibitem{Chen:2022cyw}
S.~Chen and Y.~Tanizaki, \emph{{Solitonic Symmetry beyond Homotopy:
  Invertibility from Bordism and Noninvertibility from Topological Quantum
  Field Theory}}, Phys. Rev. Lett. \textbf{131} (2023), no.~1, 011602,
  \texttt{2210.13780}.

\bibitem{Bashmakov:2022uek}
V.~Bashmakov, M.~Del~Zotto, A.~Hasan, and J.~Kaidi, \emph{{Non-invertible
  symmetries of class S theories}}, JHEP \textbf{05} (2023), 225,
  \texttt{2211.05138}.

\bibitem{Karasik:2022kkq}
A.~Karasik, \emph{{On anomalies and gauging of U(1) non-invertible symmetries
  in 4d QED}}, SciPost Phys. \textbf{15} (2023), no.~1, 002,
  \texttt{2211.05802}.

\bibitem{GarciaEtxebarria:2022jky}
I.~n. Garc\'\i{}a~Etxebarria and N.~Iqbal, \emph{{A Goldstone theorem for
  continuous non-invertible symmetries}}, JHEP \textbf{09} (2023), 145,
  \texttt{2211.09570}.

\bibitem{Choi:2022fgx}
Y.~Choi, H.~T. Lam, and S.-H. Shao, \emph{{Non-invertible Gauss law and
  axions}}, JHEP \textbf{09} (2023), 067,  \texttt{2212.04499}.

\bibitem{Yokokura:2022alv}
R.~Yokokura, \emph{{Non-invertible symmetries in axion electrodynamics}},
  (2022),  \texttt{2212.05001}.

\bibitem{Bhardwaj:2022maz}
L.~Bhardwaj, L.~E. Bottini, S.~Schafer-Nameki, and A.~Tiwari,
  \emph{{Non-Invertible Symmetry Webs}},  (2022),  \texttt{2212.06842}.

\bibitem{Bartsch:2022ytj}
T.~Bartsch, M.~Bullimore, A.~E.~V. Ferrari, and J.~Pearson,
  \emph{{Non-invertible Symmetries and Higher Representation Theory II}},
  (2022),  \texttt{2212.07393}.

\bibitem{Kaidi:2023maf}
J.~Kaidi, E.~Nardoni, G.~Zafrir, and Y.~Zheng, \emph{{Symmetry TFTs and
  Anomalies of Non-Invertible Symmetries}},  (2023),  \texttt{2301.07112}.

\bibitem{Lin:2023uvm}
Y.-H. Lin and S.-H. Shao, \emph{{Bootstrapping noninvertible symmetries}},
  Phys. Rev. D \textbf{107} (2023), no.~12, 125025,  \texttt{2302.13900}.

\bibitem{Chen:2023czk}
S.~Chen and Y.~Tanizaki, \emph{{Solitonic symmetry as non-invertible symmetry:
  cohomology theories with TQFT coefficients}},  (2023),  \texttt{2307.00939}.

\bibitem{McGreevy:2022oyu}
J.~McGreevy, \emph{{Generalized Symmetries in Condensed Matter}},  (2022),
  \texttt{2204.03045}.

\bibitem{Gomes:2023ahz}
P.~R.~S. Gomes, \emph{{An introduction to higher-form symmetries}}, SciPost
  Phys. Lect. Notes \textbf{74} (2023), 1,  \texttt{2303.01817}.

\bibitem{Schafer-Nameki:2023jdn}
S.~Schafer-Nameki, \emph{{ICTP Lectures on (Non-)Invertible Generalized
  Symmetries}},  (2023),  \texttt{2305.18296}.

\bibitem{Brennan:2023mmt}
T.~D. Brennan and S.~Hong, \emph{{Introduction to Generalized Global Symmetries
  in QFT and Particle Physics}},  (2023),  \texttt{2306.00912}.

\bibitem{Bhardwaj:2023kri}
L.~Bhardwaj, L.~E. Bottini, L.~Fraser-Taliente, L.~Gladden, D.~S.~W. Gould,
  A.~Platschorre, and H.~Tillim, \emph{{Lectures on Generalized Symmetries}},
  (2023),  \texttt{2307.07547}.

\bibitem{Luo:2023ive}
R.~Luo, Q.-R. Wang, and Y.-N. Wang, \emph{{Lecture Notes on Generalized
  Symmetries and Applications}},  (2023),  \texttt{2307.09215}.

\bibitem{Shao:2023gho}
S.-H. Shao, \emph{{What's Done Cannot Be Undone: TASI Lectures on
  Non-Invertible Symmetry}},  (2023),  \texttt{2308.00747}.

\bibitem{Iqbal:2021rkn}
N.~Iqbal and J.~McGreevy, \emph{{Mean string field theory: Landau-Ginzburg
  theory for 1-form symmetries}}, SciPost Phys. \textbf{13} (2022), 114,
  \texttt{2106.12610}.

\bibitem{Migdal:1983qrz}
A.~A. Migdal, \emph{{Loop Equations and 1/N Expansion}}, Phys. Rept.
  \textbf{102} (1983), 199--290.

\bibitem{Makeenko:1980vm}
Y.~Makeenko and A.~A. Migdal, \emph{{Quantum Chromodynamics as Dynamics of
  Loops}}, Sov. J. Nucl. Phys. \textbf{32} (1980), 431.

\bibitem{Polyakov:1980ca}
A.~M. Polyakov, \emph{{Gauge Fields as Rings of Glue}}, Nucl. Phys. B
  \textbf{164} (1980), 171--188.

\bibitem{Nambu:1961tp}
Y.~Nambu and G.~Jona-Lasinio, \emph{{Dynamical Model of Elementary Particles
  Based on an Analogy with Superconductivity. 1.}}, Phys. Rev. \textbf{122}
  (1961), 345--358.

\bibitem{Goldstone:1961eq}
J.~Goldstone, \emph{{Field Theories with Superconductor Solutions}}, Nuovo Cim.
  \textbf{19} (1961), 154--164.

\bibitem{Goldstone:1962es}
J.~Goldstone, A.~Salam, and S.~Weinberg, \emph{{Broken Symmetries}}, Phys. Rev.
  \textbf{127} (1962), 965--970.

\bibitem{Lake:2018dqm}
E.~Lake, \emph{{Higher-form symmetries and spontaneous symmetry breaking}},
  (2018),  \texttt{1802.07747}.

\bibitem{Hofman:2018lfz}
D.~M. Hofman and N.~Iqbal, \emph{{Goldstone modes and photonization for higher
  form symmetries}}, SciPost Phys. \textbf{6} (2019), no.~1, 006,
  \texttt{1802.09512}.

\bibitem{Hidaka:2019mfm}
Y.~Hidaka, M.~Nitta, and R.~Yokokura, \emph{{Emergent discrete 3-form symmetry
  and domain walls}}, Phys. Lett. B \textbf{803} (2020), 135290,
  \texttt{1912.02782}.

\bibitem{Makeenko:2002uj}
Y.~Makeenko, \emph{{Methods of contemporary gauge theory}}, Cambridge
  Monographs on Mathematical Physics, Cambridge University Press, 11 2005.

\bibitem{KAWAI199293}
H.~Kawai, \emph{Quantum gravity and random surfaces}, Nuclear Physics B -
  Proceedings Supplements \textbf{26} (1992), 93--110.

\bibitem{Leigh:1989jq}
R.~G. Leigh, \emph{{Dirac-Born-Infeld Action from Dirichlet Sigma Model}}, Mod.
  Phys. Lett. A \textbf{4} (1989), 2767.

\bibitem{Hidaka:2020ucc}
Y.~Hidaka, Y.~Hirono, and R.~Yokokura, \emph{{Counting Nambu-Goldstone Modes of
  Higher-Form Global Symmetries}}, Phys. Rev. Lett. \textbf{126} (2021), no.~7,
  071601,  \texttt{2007.15901}.

\bibitem{Villain:1977}
{Villain, J.}, \emph{A magnetic analogue of stereoisomerism : application to
  helimagnetism in two dimensions}, J. Phys. France \textbf{38} (1977), no.~4,
  385--391.

\bibitem{Nielsen:1975hm}
H.~B. Nielsen and S.~Chadha, \emph{{On How to Count Goldstone Bosons}}, Nucl.
  Phys. B \textbf{105} (1976), 445--453.

\bibitem{Schafer:2001bq}
T.~Sch\"afer, D.~T. Son, M.~A. Stephanov, D.~Toublan, and J.~J.~M.
  Verbaarschot, \emph{{Kaon condensation and Goldstone's theorem}}, Phys. Lett.
  B \textbf{522} (2001), 67--75,  \texttt{hep-ph/0108210}.

\bibitem{Miransky:2001tw}
V.~A. Miransky and I.~A. Shovkovy, \emph{{Spontaneous symmetry breaking with
  abnormal number of Nambu-Goldstone bosons and kaon condensate}}, Phys. Rev.
  Lett. \textbf{88} (2002), 111601,  \texttt{hep-ph/0108178}.

\bibitem{Nambu:2004yia}
Y.~Nambu, \emph{{Spontaneous Breaking of Lie and Current Algebras}}, J.
  Statist. Phys. \textbf{115} (2004), no.~1/2, 7--17.

\bibitem{Watanabe:2011ec}
H.~Watanabe and T.~Brauner, \emph{{On the number of Nambu-Goldstone bosons and
  its relation to charge densities}}, Phys. Rev. D \textbf{84} (2011), 125013,
  \texttt{1109.6327}.

\bibitem{Watanabe:2012hr}
H.~Watanabe and H.~Murayama, \emph{{Unified Description of Nambu-Goldstone
  Bosons without Lorentz Invariance}}, Phys. Rev. Lett. \textbf{108} (2012),
  251602,  \texttt{1203.0609}.

\bibitem{Hidaka:2012ym}
Y.~Hidaka, \emph{{Counting rule for Nambu-Goldstone modes in nonrelativistic
  systems}}, Phys. Rev. Lett. \textbf{110} (2013), no.~9, 091601,
  \texttt{1203.1494}.

\bibitem{Watanabe:2014fva}
H.~Watanabe and H.~Murayama, \emph{{Effective Lagrangian for Nonrelativistic
  Systems}}, Phys. Rev. X \textbf{4} (2014), no.~3, 031057,
  \texttt{1402.7066}.

\bibitem{Hayata:2014yga}
T.~Hayata and Y.~Hidaka, \emph{{Dispersion relations of Nambu-Goldstone modes
  at finite temperature and density}}, Phys. Rev. D \textbf{91} (2015), 056006,
   \texttt{1406.6271}.

\end{thebibliography}

\end{document}